\documentclass[AEJ]{AEJ}
\usepackage{lscape}
\usepackage{setspace}
\usepackage{graphicx}
\usepackage{color}
\usepackage{enumerate}
\usepackage{multirow}
\usepackage{comment}
\usepackage[]{authblk}
\usepackage[round]{natbib}
\usepackage{pdflscape}
\usepackage[figuresleft]{rotating}
\usepackage{adjustbox}
\usepackage{graphics}
\usepackage{verbatim}
\usepackage{amsmath, amssymb}
\usepackage{bbm}
\usepackage{verbatim}
\usepackage{varwidth}
\usepackage[english]{babel}
\usepackage[utf8]{inputenc}
\usepackage{xr}
\usepackage{hyperref}
\usepackage{color,soul}
\usepackage{tabularx}
\usepackage{booktabs}
\usepackage{siunitx}
\usepackage{color, colortbl}
\usepackage{threeparttable}
\usepackage{dirtytalk}
\definecolor{Gray}{gray}{0.9}

\hypersetup{
    colorlinks=true,
    citecolor=red,
    linkcolor=blue,
    filecolor=magenta,      
    urlcolor=blue
}

\draftSpacing{2}

\begin{document}

%\DOI{10.1287/xxxx.0000.0000}%


\title{Gender Bias and Property Taxes}
\shortTitle{Gender Bias and Property Taxes}
\author{Gordon Burtch\thanks{We would like to thank Emma Wiles, Jetson Leder-Luis, Dokyun Lee, Bin Gu, Patricia Cortés, and Zoe Cullen for helpful comments and suggestions, as well as seminar participants at Southern Methodist University, George Washington University, U Mass-Amherst, and Boston University. Joanna Jia, Heetal Binwani, and Tiffany Zhang provided superb research assistance.}\footnote{Boston University -- gburtch@bu.edu}, \& Alejandro Zentner\footnote{UT Dallas -- azentner@utdallas.edu}}

\date{\today}
%\pubMonth{Month}
%\pubYear{Year}
%\pubVolume{Vol}
%\pubIssue{Issue}

\begin{abstract}
Gender bias distorts the economic behavior and outcomes of women and households. We investigate gender biases in property taxes. We analyze records of more than 100,000 property tax appeal hearings and more than 2.7 years of associated audio recordings, considering how panelist and appellant genders associate with hearing outcomes. We first observe that female appellants fare systematically worse than male appellants in their hearings. Second, we show that, whereas male appellants' hearing outcomes do not vary meaningfully with the gender composition of the panel they face, those of female appellants' \textit{do}, such that female appellants obtain systematically lesser (greater) reductions to their home values when facing female (male) panelists. Employing a multi-modal large language model (M-LLM), we next construct measures of participant behavior and tone from hearing audio recordings. We observe markedly different behaviors between male and female appellants and, in the case of male appellants, we find that these differences \emph{also} depend on the gender of the panelists they face (e.g., male appellants appear to behave systematically more aggressively towards female panelists). In contrast, the behavior of female appellants remains relatively constant, regardless of their panel's gender. Finally, we show that female appellants continue to fare worse in front of female panels, even when we condition upon the appellant's in-hearing behavior and tone. Our results are thus consistent with the idea that gender biases are driven, at least in part, by unvoiced perceptions among ARB panelists. Our study documents the presence of gender biases in property appraisal appeal hearings and highlights the potential value of generative AI for analyzing large-scale, unstructured administrative data. \\ \\
\end{abstract}


\maketitle

%\begin{center}
%    \textbf{Note:} Please do not circulate or cite without the authors' permission.
%\end{center}

\clearpage
\section{Introduction}
\label{sec:introduction}

Gender disparities in evaluations, opportunities, and outcomes are well documented across a wide range of economically significant domains, particularly in contexts that involve subjective judgment. A growing body of research demonstrates that, in many cases, gender concordance or discordance—congruence or incongruence of gender between evaluators and those being evaluated—can drive these disparities, influencing outcomes in administrative processes, the labor market, education, healthcare, and criminal justice settings \citep[e.g.,][]{cullen2023old,cabral2024gender,takeshita2020association}. These effects matter because they suggest that the personal characteristics of evaluators and evaluated individuals may shape critical decisions, often leading to systematic bias.

We examine gender concordance effects in the economically important context of property taxation, specifically in the administrative process related to property appraisal appeals \citep[e.g.,][]{avenancio2022assessment, nathan2020my}. Property taxes represent the second-largest source of tax revenue in the United States, contributing an estimated \$547 billion in 2019, more than doubling revenue from corporate income tax \citep{giaccobasso2022my}. Property taxes are particularly important for sub-national governments, which rely on revenue from property taxes to fund essential services such as schools, police, and parks. 

In every U.S. county that we are aware of, households can file an appeal over their property appraisal, seeking to reduce their tax burden. Households can appeal their appraised value if they believe it exceeds the market value, if comparable properties have been appraised at lower values leading to an unfair tax burden, or if there are errors in the county's information used to assess the property. 

County Appraisal Districts (CADs) in the U.S. rely on data and statistical models to estimate a home's market value. However, these estimates can vary significantly, as evidenced by the wide differences in appraised values assigned to the same home by large companies like Zillow or Redfin. Households leverage this inherent subjectivity in the appraisal process to make the case for an appeal, seeking to reduce their tax burden. In many cases, appeals in our focal state of Texas proceed to a formal, live hearing, wherein a homeowner or their representative presents the case for their appeal in front of a panel of appraisal review board (ARB) members.

We study data on property appraisal appeal hearings obtained from Harris County, Texas (the Houston area) and associated audio recordings. Exploiting exogenous variation in the composition of ARB hearing panels attributable to the fact that panelists are randomly assigned to hearings, we examine how appeal outcomes for female and male appellants may differ depending on the gender composition of their hearing panel. We focus on gender concordance between appellants and panelists, documenting several notable findings. We calculate panelist-specific savings adjudication rates and document significant heterogeneity, beyond what could be explained by chance. Further, we calculate estimates of panelist-specific differences in savings adjudication rates by appellant gender, reflective of anti-female (pro-male) bias, and find that female panelists are systematically less likely to adjudicate in favor of a female appellant as compared to a male appellant. Next, undertaking more formal regression analyses, we find consistent results; we show that female appellants are systematically less likely to be awarded reductions in their home value and that this is particularly true when their ARB panel is comprised of females. The effects we report are substantial; we estimate that female appellants are approximately 4.2 p.p. less likely to obtain any reduction to their home value when faced with an all-female panel as compared to an all-male panel. Further, we estimate that, when facing an all-female (versus all-male) panel, female appellants obtain an approximate 33\% smaller reduction to their initial home appraisal, on average. Given the average home appraisal value reduction across all hearings is approximately \$20,200, the gender concordance effect we observe translates to a roughly \$6,700 difference in final appraised value and an approximate \$140 difference in annual property taxes owed (the average home in Harris County presently pays a tax rate of 2.13\%). 

Next, leveraging unstructured audio recordings associated with more than 80,000 of these hearings, i.e., more than 2.7 years of audio, we employ a multi-modal large language model (M-LMM), Gemini 1.5 Flash \citep{gemini2024}, to annotate appellant and panelist behaviors and tones in each hearing -- a task that would be prohibitively costly employing human analysts. Our goal is to understand why the observed gender differentials might arise. Examining differences in the prevalence of behaviors and tones across hearings by appellant and panel chair gender, we first observe that appellants exhibit much more variation in their tone and behavior than panelists. This can be explained, to some extent, by the fact that panelists receive formal training and are aware that their behavior and statements may be subsequently reviewed in court should the appellant decide to appeal their decision further.

Considering appellant behaviors and tones, we find that female appellants differ, in general, from male appellants in several respects. For example, females are coded as being systematically less likely to 'sound confident,' they are less likely to present formal evidence, and they are less likely to disagree verbally with a panelist.\footnote{We considered the possibility that Gemini may be biased in its evaluation of men versus women by comparing Gemini's annotations for a random sample of 75 audio recordings with annotations obtained from a large sample of human annotators on Prolific. Notably, we demonstrate substantial agreement between Gemini and humans.} Most important, of course, are behaviors or tones that differ systematically with the gender of ARB panelists, as those behaviors might help explain differential hearing outcomes. And, indeed, we observe that some appellant behaviors and tones \textit{do} differ in ways that depend on the panel chair's gender. 

For example, male appellants behave systematically more aggressively toward female panelists. We observe that male appellants are systematically more likely to 'sound hostile', to raise their voice, to sound frustrated, to sound annoyed, to sound defensive, and to interrupt when a panelist is speaking, specifically in front of a female panel chair, as compared to a male panel chair, and as compared to female appellants, broadly. At the same time, we observe that female appellants' are more likely to sound confident in front of a female panel chair, they are more likely to ask a panelist to repeat themselves, and they are more likely to ask a panelist to 'speak up'. However, in each case where female appellants' behavior differs with ARB panel chair gender, that difference is relatively small in magnitude. 

Of course, behavioral differences will not necessarily lead to differential hearing outcomes if social norms regarding expected behavior for males and females differ. This leads to our final, and perhaps most notable finding. A unique aspect of our study and dataset is that we can separate the influence of changes in appellant behavior from the influence of differences in panelist perceptions. By controlling for appellant behavior, we can infer the role of panelist perceptions. That we observe males behaving much more aggressively toward female panels, yet experiencing systematically more positive hearing outcomes already suggests a clear role for differences in panelist perceptions. Examining these effects more explicitly, we find that females remain systematically less likely to obtain home value reductions from female panelists, conditional on the appellant's in-hearing behavior and tone. The magnitude of the conditional effects leads us to conclude that the differential hearing outcomes are largely attributable to unvoiced perceptions or beliefs among ARB panelists.

Our findings have several important implications. The negative, and female-appellant specific gender concordance effect that we document reveals previously unrecognized biases that have the potential to manifest more broadly in fundamental administrative processes across the United States. Property tax appeals directly impact individuals’ financial burdens and the amount they contribute toward the provision of public goods and services, e.g., public schools. Fairness in these hearings is essential for maintaining equity and trust in the tax system. Our findings are most likely to extend to other settings where evaluations are undertaken by peers or panels, such as academic peer reviews, corporate board decisions, judicial proceedings, administrative processes (e.g., driver’s license issuance), and private decision-making contexts (e.g., loan requests). Our study underscores the need for policymakers to be aware of potential gender biases that may arise in these settings, such that they might consider structural or procedural adjustments, to mitigate their manifestation. Our work also highlights how gender differentials and biases may be evaluated and interrogated by employing large-scale administrative data, including unstructured records of procedural hearings.  

\section{Related Work}
\label{sec:lit-review}

A great deal of literature has considered the impact of demographic concordance on evaluative outcomes. Much of that work has documented the role of gender concordance, showing that shared gender relates positively to evaluation outcomes, from patient evaluations of physicians \citep{lau2021does,cabral2024gender}, to hiring and compensation \citep{coffman2021role,cullen2023old}, to judicial decisions in the courtroom \citep{boyd2017effects,knepper2018shadow}. 

The mechanisms behind gender concordance effects are often nuanced and highly context-specific. Demographic concordance can facilitate more positive evaluations due to improved communication and empathy. However, literature dealing with social identity and cultural stereotypes points to a variety of context-specific reasons why gender concordance may benefit an evaluated individual, particularly in terms of gender stereotypes and gender roles. Particularly relevant to our study context, past literature has documented that oversight of household finances is often characterized by a strong gender stereotype, namely the expectation that males may take the lead \citep{bordalo2016stereotypes,brock2023discriminatory}. This gender stereotype may cause women to exert less influence in property tax appeal hearings, as women may form negative self-assessments of their competence in the domain \citep{coffman2014evidence}. This, in turn, may lead to a positive gender concordance effect if, for example, evaluating women actively seek to overcome the gender stereotype by engaging in activist choice homophily \citep{greenberg2017activist}, a behavior that has been observed in the context of entrepreneurial finance, with female funders actively seeking to support female technology entrepreneurs as a means of combating the marginalization of women in tech. 

There is also real potential that gender concordance will lead to negative outcomes for women. Research has found that gender dynamics can have a strong influence on negotiation outcomes, with women facing greater expectations and challenges than men \citep{bowles2007social,recalde2023gender,cortes2024gender}. Literature on social identity suggests that in-group members may judge each other more harshly in settings where they may be concerned about possible accusations of favoritism \citep{tajfel1979integrative}. This idea has been argued to contribute to the ‘queen bee’ syndrome, wherein women who ascend to leadership positions in male-dominated workplaces may distance themselves from other, more junior women \citep{ely1994effects}.   

Of course, beyond the broad literature dealing with gender concordance effects, our paper relates and contributes to the literature studying what motivates households to appeal their property taxes. In this literature, \citet{nathan2020my} show that both expected tax savings and filing frictions play significant roles, \citet{giaccobasso2022my} show that households are less likely to appeal if they believe that the government services funded by those taxes will provide more significant personal benefit to them, and \cite{Nathan2024} show that households are willing to pay more in taxes if they believe other households are contributing their fair share. Some papers in this literature study property tax appeals heterogeneity by race, gender, and partisanship \citep{  avenancio2022assessment,Nathan2022, nathan2020my, AndrewJustin2023}.\footnote{For other studies on property taxes, see, for example, \citet{cabral2012hated} and \citet{KRESCH2023}} While these studies focus on what motivates the initial decision to appeal, our focus is on biases that affect the process itself, at the ARB hearings stage. Of particular relevance to our study, \citet{AndrewJustin2023} describes the results of an experiment that involved sending emails to county appraisers, posing as homeowners, and randomly varying the homeowner's gender. Those authors find that female appraisers are less likely to answer emails from a female homeowner. Although \citet{AndrewJustin2023} does not focus on appeal hearings, as we do, their result is initially suggestive of possible negative gender concordance effects in the property tax context. It should be noted, however, that appraisers play a rather marginal role in hearing outcomes in our setting; despite appraisers' presence at the hearings, final decisions are ultimately rendered by the ARB panel.

Finally, our paper contributes to the literature on unstructured data and open-ended survey questions \citep[e.g.,][]{Gentzkow2010,Gentzkow2019,Stantcheva2020,Stantcheva2023}. Large natural language models offer significant potential to expand the use of unstructured data. We advance this area of research by using these generative AI tools to process 2.7 years of audio recordings, a virtually impossible task for humans.

\section{Methods}
\label{sec:data-and-methods}

We divide this section into several parts. In the first subsection, we introduce our study context. In the second subsection, we present our data and measures. We then discuss identification and our econometric specification, followed by our approach to processing the unstructured audio data. 

\subsection{Study Context}
\label{sec:context}

Our study leverages data related to property appraisal appeal hearings conducted in Harris County, Texas. Harris County is the third largest county in the United States, with approximately 4.2 million residents as of 2024. Harris County residents are ethnically diverse\footnote{https://www.houstonchronicle.com/projects/2023/houston-population-ethnicity/} and politically moderate,\footnote{https://www.bakerinstitute.org/research/texas-counties} making it an ideal context for our analysis. 

In Texas, County appraisal districts () appraise all homes within their counties and post these appraised values on their websites, sometimes also sending a \say{Notice of Appraised Value} by mail. These values are typically available by April 15 every year, and households have one month to file an appeal if they disagree with the proposed appraised value. 

Protests can be filed for various reasons, including a) market value discrepancies, i.e., homeowners may argue the assessed value is higher than the market value of similar properties recently sold in the county, which often involves comparing the price of their property to the prices of comparable homes from recent sales, known as \say{comps,} b) unequal appraisal, i.e., the homeowner may argue that their property’s proposed taxable value is disproportionately high compared to the taxable value of comparable properties within the same district or neighborhood, potentially signaling unequal appraisal practices leading to unfair taxation, and c) other reasons, such as denied exemptions, errors in the public records (e.g., an incorrect number of bedrooms, bathrooms, or square footage), etc. 

Homeowners can file an appeal independently, or hire an agent to appeal on their behalf. After an appeal is filed, CADs may offer a settlement, usually via email. If the CADs do not offer a settlement or homeowners do not accept the proposed settlement, the appeal moves to the , a group of citizens appointed by the local administrative district judge to resolve disputes between property owners and the CADs. 

Appraisal District Boards in Texas, such as the Harris County Appraisal District (HCAD) in the Houston area, are political subdivisions of the Texas Comptroller of Public Accounts, and homeowners in every county in Texas can appeal using the same form provided by the state. For this reason, tax appeals operate very similarly across all 241 counties in Texas.

In Texas, as in many other states, property tax protests may be filed by or on behalf of any resident. Any given protest, should it proceed to a formal hearing, is adjudicated by a panel of ARB members drawn (randomly) from a pool of several hundred. To serve as a member of the ARB in Harris County (the focus of this study), an individual must apply and undergo an interview with a local district judge or their representative. Applicants for the position must meet several requirements. Applicants must have been a county resident for at least two years, and must not hold any conflicts of interest. An applicant must also be willing and able to attend ARB hearings on any given weekday or Saturday and must be willing and able to attend hearings conducted online or in person, at the discretion of the appellant.\footnote{Official requirements are detailed on the \href{https://hcad.seamlessdocs.com/f/rjr3ak3eae0r}{Harris County ARB panelist application form}.} Importantly, appellants are not made aware of the identity of their ARB panel in advance of the hearing. Hence, appellants have no ability to self-select, e.g., deciding whether a husband or wife might attend, based on their panel's composition.

Our structured data pertain to all property tax protests that proceeded to a formal hearing with the Harris County ARB, from 2013 through mid-2023, where the appellant appearing at the hearing was an individual homeowner.\footnote{Generally, the attendee that appears at a given hearing may be a property owner or a representative of a professional agency that files property tax protests on behalf of their clients. Our study focuses on individuals who represent their own property because protest agents are professionals. As a result, professional agents are likely to experience very different outcomes. Protest agents frequently represent large volumes of clients in front of the ARB; hence, they have more experience with the protest process, and they are also more likely to be familiar with ARB panelists through the course of their work.} The initial sample includes a total of 114,515 protest hearings. Additionally, we consider audio recordings associated with 80,197 of these hearings.

\subsection{Data and Measures}
\label{sec:data-and-methods:dscstats}

\subsubsection{Structured Data}

Each protest hearing record includes information on the protested property value assessment, as well as the revised assessment (if any) that emerged as an outcome of the hearing, i.e., the adjusted property value. We consider the difference between these two measures. Specifically, we construct a binary indicator of $I(Saved)$ and a continuous, non-negative measure of the percentage saved, i.e., $Pct Saved$, where the latter reflects the percentage decline from the original assessed value. The former value thus reflects savings that appellants receive at the extensive margin, while the latter reflects savings at both the extensive and intensive margin. In approximately 243 cases, the final assessed property value exceeds the original property value assessment. This is a rare occurrence, which likely reflects that pieces of information came to light during the hearing that was detrimental to the appellant's case, leading the ARB to realize that the initial assessment was too low. We omit these hearings from our analyses. 

Each protest hearing record also includes the names of the ARB panelists who participated in the hearing, the appraiser, and the individuals who attended the formal hearing as the appellant(s). We parse the first and last names of each party from the hearing record. We then impute the gender of each based on their parsed first name.\footnote{This imputation is based on Social Security Administration (SSA) data \citep{kaplan2021predictrace}. It should be noted that multiple parties may attend the hearing representing the property in question. In such cases, we gender code the first name that is listed. We focus our analysis on the first appellant listed because, unlike the other names of ARB panelists and the appraiser, which are standardized and consistently reported -- suggesting these values are populated from an IT system -- all appellant names are reported together in a single field, and the formatting is highly inconsistent -- presumably reflecting manual data entry. We elaborate on the implied measurement error later and the steps we take to address that error leveraging gender detected based on voices in our audio recordings. Gender inferences are made based on empirical probabilities derived from the SSA data. For example, if a particular first name is held by 100 people in the United States SSA records, and 98 of those people were identified as female, then we would infer a 98\% probability that the individual in our sample is female. Our approach employs a majority threshold in assigning gender. However, our results are robust to using stricter thresholds, e.g., 80\%, 90\%, 95\%, and indeed they grow stronger.} In turn, we use this gender information to construct an indicator of appellant gender, as well as a count-based measure of the number of $FemalePanelists$ that were present. 

Each protest record includes a hearing case description field, populated with categorical values conveying details of the hearing, including, most notably, the medium via which the hearing took place. Beginning in 2020, due to the COVID-19 pandemic, Harris County began allowing protesters to elect that their hearing takes place online, via WebEx or Zoom. This practice was continued with the decline of the pandemic and remains in place today. A value of VRTL indicates that a hearing was conducted online, via either WebEx or Zoom. This information is used to construct a binary indicator of the presence of this code in the hearing record, $Online$.

We also consider several variables as controls. We construct dummies reflecting the tax year in which a protest was filed. Further, we construct a time-series measure of the cumulative number of property tax protest hearings that appear in the data before the tax year of a given protest, $PriorProtests$, as well as an indicator of whether the most recent protest, should it exist, resulted in savings for the appellant, $RecentSavings$. Finally, we construct two dummy indicators reflecting the basis or rationale of the protest, whether the appellant was arguing a market value discrepancy, and whether they were arguing an unequal appraisal value. All other arguments are taken as the reference category.

Lastly, it is important to note that, beginning in 2022, homeowners began to receive the option to present their case in front of a one-member or three-member panel. Accordingly, a subset of the hearing records document only one ARB panelist. We will consider estimations that model outcomes are a function of only the gender of the ARB panel chair. For hearings conducted before a three-member panel, we will also consider estimations modeling outcomes as a function of the count of female (vs. male) panelists. 

Table~\ref{tab:desc} reports descriptive statistics for the variables based on our structured dataset of 114,515 hearings.\footnote{Note that this sample reflects the subset of all structured hearing recordings, namely the subset for which we successfully i) employ regular expressions to parse a first name and ii) predict gender for all hearing attendees based on those first names, including all ARB panelists, appraisers, and the first appellant listed. Should we fail to parse a name that can be matched to SSA data, we ultimately omit the hearing recording. As a result, we omitted approximately 59,000 hearing records from the original sample where the gender prediction failed.} We see that the mean assessed home value in this sample is approximately \$407,000, that approximately 20\% of hearings have been conducted online, and that approximately 65\% of hearings result in some tax savings for the appellant, translating to an average reduction of roughly \$23,000 to the assessed home value. Further, concerning gender, we see that roughly 34\% of hearing appellants are female, that approximately 55\% of ARB panels are chaired by a female, and that the three-member panels include 1.55 females on average, suggesting strong gender balance in the ARB panelist pool. 

Before we proceed further, we can assess the ARB's claim that panelists are, in fact, randomly assigned to hearings. We can evaluate this via a balance check, assessing covariate imbalance in hearing characteristics between ARB panels chaired by a male and female panelist. We assess covariate imbalance employing standardized mean differences (SMD), i.e., the difference in means in units of the pooled standard deviation \citep{rosenbaum1985constructing}.\footnote{Unlike t-tests, the standardized mean difference is not influenced by sample size. The standardized mean difference also enables a comparison of the relative balance across variables measured in different units.} The results of the balance test are reported in Table~\ref{tab:bal_check}. All variables exhibit an SMD well below 0.10, a common threshold for practically meaningful imbalance in the literature \citep{austin2009balance}.

\begin{table}[htpb!]
\small
\centering
\caption{Descriptive Statistics: Structured Data}
\label{tab:desc}
\begin{tabular}{lcc}
\toprule
\textbf{Variable} & \textbf{Mean} & \textbf{SD}  \\ 
\hline
Female Appellant      & 0.343   & 0.475        \\ 
Female Panel Chair     & 0.552   & 0.497   \\ 
Count Female Panelists     & 1.574   & 0.949        \\ 
Any Saved    & 0.651   & 0.477         \\ 
Assessed Home-value Decrease           & 22,946.383  & 63,540.795  \\ 
Online               & 0.204   & 0.403       \\ 
Home Value      & 407,010.893 & 537,843.339  \\ 
Tax Year             & 2018.120	& 3.200       \\ 
Protests Prior       & 0.621   & 1.122      \\ 
Recent Savings  &  0.241 & 0.428  \\
\hline
\end{tabular}
\end{table}

\begin{table}[htpb!]
\small
\centering
\caption{Standardized Mean Differences in Hearing Characteristics By Panel Chair Gender}
\begin{tabular}{lccccc}
\toprule
 & \multicolumn{2}{c}{\textbf{Male}} & \multicolumn{2}{c}{\textbf{Female}}  \\
\textbf{Variable} & \textbf{Mean} & \textbf{SD} & \textbf{Mean} & \textbf{SD} & \textbf{SMD} \\
\midrule
Appellant Female  & 0.346 & (0.476) & 0.341  & (0.474) & -0.0114 \\
Log(Home Value)  & 12.5 & (0.851)  & 12.5 & (0.857)  & 0.0795 \\
Prior Protests   & 0.604 & (1.11)  & 0.634 & (1.13)  & 0.0274 \\
Recent Savings & 0.237 & (0.425) & 0.245 & (0.430) & 0.0195 \\
Online          & 0.196 & (0.397) & 0.211 & (0.408) & 0.0394 \\
Tax Year         & 2018 & (3.24)   & 2018 & (3.17)   & 0.0726 \\
\bottomrule
\end{tabular}
\label{tab:bal_check}
\end{table}

\subsubsection{Audio Data}

In addition to our structured sample, we obtained audio recordings associated with thousands of hearings. This data is valuable because it allows us to construct measures of what took place during hearings, with the potential to shed light on the sources of any bias we might document in hearing outcomes, based on our structured dataset. Additionally, we can leverage the recordings to explore any possible role of measurement error around appellant gender, considering alternative measures based on an appellant's voice. Given the ARB indicates a typical hearing should last about 15 minutes, we omit recordings that are excessively long or short in duration (more than 45 minutes or less than 3 minutes). We then match the resulting recordings to structured hearing records in our sample, achieving a match for 80,197 hearings, our working sample. The total audio duration in this sample is approximately 2.7 years (more than 23,650 hours). Listening to each recording and coding its characteristics manually would be infeasible. Accordingly, we process the audio recordings employing Google's Gemini 1.5 Flash, a multi-modal large language model (LLM) capable of processing lengthy audio input. For each audio recording, we ask Gemini to \say{listen} to the file and respond to several questions about the behavior and tone of the 'appellant who speaks the most' and the 'panelist who speaks the most', replying with a consistently (JSON) formatted response. 

We use the annotations from Gemini to i) construct an alternative, complementary indicator of appellant gender, ii) explore possible differences in how appellants act depending on their gender and the gender of ARB panelists, iii) examine how different behaviors and characteristics relate to hearing outcomes, depending on appellant and ARB panelist gender, and iv) whether hearing outcomes differ by appellant and ARB panelist gender, conditional on the in-hearing behavior and tone of appellants. It should be noted that we coded features that capture both behavior and tone of the 'appellant who speaks the most' and of the 'panelist who speaks the most'. This is important because the appellant who speaks most might not be the first appellant listed in the structured hearing record. Further, the panelist who speaks the most might not be the chair. It should also be noted that we did not attempt to code the gender of all three ARB panelists in the case of three-member panels, because the number of people in the room would be rather numerous, and the task would require a combination of speaker identification and gender prediction, a rather complex task, one for which we would lack confidence in Gemini Flash's ability.

Table~\ref{tab:audio_desc} reports descriptive statistics for the variables derived from our audio dataset. We provide more detail about the specific behaviors and tones we code from the audio files in the Appendix. A particularly notable pattern we observe is very little variation among the tonal and behavioral measures exhibited by the ARB panelist. Panelists are almost always coded as sounding confident, they never sound angry or frustrated, etc. By contrast, appellants exhibit wide variation in behavior and tone. This difference is likely driven by the fact that ARB panelists follow a well-defined script and procedure during the hearings. Moreover, ARB panelists receive formal training on how to behave and what statements they should or should not make during hearings. This is particularly important because ARB panelists are generally aware that the outcomes of protest hearings may appealed in court. 

\begin{table}[ht]
\small
\centering
\caption{Descriptive Statistics: Audio Data}
\label{tab:audio_desc}
\begin{tabular}{lcc}
\toprule
\textbf{Variable} & \textbf{Mean} & \textbf{SD}  \\
\midrule
Appellant Presents Evidence & 0.589 & 0.492 \\
Appellant Disagrees with Panelist & 0.759 & 0.428 \\
Appellant Raises Voice & 0.146 & 0.353 \\
Appellant Asks Questions & 0.718 & 0.450 \\
Appellant Interrupts Panelist & 0.335 & 0.472 \\
Appellant Asks Panelist to Speak Louder & 0.065 & 0.246 \\
Appellant Asks Panelist to Repeat & 0.348 & 0.476 \\
Appellant Sounds Confident & 0.896 & 0.305  \\
Appellant Sounds Hostile & 0.111 & 0.314  \\
Appellant Sounds Frustrated & 0.809 & 0.393  \\
Appellant Sounds Annoyed & 0.769 & 0.422 \\
Appellant Sounds Defensive & 0.244 & 0.429  \\
Appellant Sounds Angry & 0.110 & 0.312 \\
Appellant Sounds Nervous & 0.023 & 0.149  \\
Panelist Disagrees with Appellant & 0.327 & 0.469 \\
Panelist Raises Voice & 0.000 & 0.016  \\
Panelist Asks Questions & 0.845 & 0.361 \\
Panelist Interrupts Appellant & 0.104 & 0.305  \\
Panelist Asks Appellant to Speak Louder & 0.019 & 0.136 \\
Panelist Asks Appellant to Repeat & 0.133 & 0.339  \\
Panelist Sounds Confident & 0.999 & 0.024  \\
Panelist Sounds Hostile & 0.000 & 0.015  \\
Panelist Sounds Frustrated & 0.005 & 0.071  \\
Panelist Sounds Annoyed & 0.004 & 0.067  \\
Panelist Sounds Defensive & 0.006 & 0.078  \\
Panelist Sounds Angry & 0.000 & 0.015 \\
Panelist Sounds Nervous & 0.000 & 0.007  \\
\bottomrule
\end{tabular}
\end{table}

\subsection{Econometric Specification}
\label{sec:data-and-methods:identification}

Our initial specification, detailed in Equation (\ref{eq:economtricmodel}), is intended to examine how homeowner and panelist gender relate to hearing outcomes, in terms of property tax savings. In the equation, hearings are indexed by \textit{i}, and tax years are indexed by \textit{t}. For measures of $HomeValueSavings$, we focus our initial estimations on a binary indicator of whether the homeowner achieves any reductions in their property taxes via a reduction in the assessed home value at the end of the hearing, $AnySaved^{g}_{i,t})$. We model that binary indicator as a function of the ARB panel chair's gender, $FemalePanelChair_{i,t}$, tax-year fixed effects, home value bucket fixed effects, where buckets reflect ranges of \$100,000, appraiser fixed effects, a measure reflecting the count of prior property tax protest hearings that appear in our sample for the same property up to that point, i.e., from 2013 forward, a binary indicator of whether the hearing is conducted online or offline, and a binary indicator of whether the most recent protest hearing, if it exists, resulted in some appraisal reduction (tax savings) for the property owner. Our key parameter of interest in this analysis is $\beta_1$, which captures any differences in hearing outcomes depending on the gender of the panel chair, conditional on case characteristics.

\begin{equation}\label{eq:economtricmodel}
\begin{aligned}
    Ho&meValueSavings_{i,t} = \vphantom{\sum_{v=2}} \\& 
    \vphantom{\sum_{v=2}} \beta_1 \cdot FemalePanelChair_{i,t} + \beta_2 \cdot PriorProtests_{i,t} \vphantom{\sum_{v=2}} + \beta_3 \cdot Online_{i,t}  \\
    & + \beta_4 \cdot RecentSavings_{i,t} \vphantom{\sum_{v=2}} + \beta_5 \cdot MarketEquity_{i,t} + \beta_6 \cdot UnequalAppraisal_{i,t} \\
    & + \sum^{V}_{v=2} \eta_v \cdot HomeValueBucket^v_{i,t} + \sum_{\tau=2014}^{2023} \lambda_\tau \cdot TaxYear^\tau_{i,t} + \sum_{a=2}^{A} \nu_a \cdot Appraiser^a_{i,t} + \epsilon_{i,t} 
\end{aligned}
\end{equation}

We also conduct estimations that replace our binary indicator of a female panel chair gender with a series of three dummies reflecting the count of female panelists (taking 0 as a reference), limiting the estimation to only those hearings that involved three-member ARB panels. Further, we report a series of estimations incorporating account (property) fixed effects to ensure our results are not a function of unobserved imbalances in the features of the properties involved. 

We next replace our binary outcome, \textit{AnySaved}, with a non-negative measure of the dollar amount saved, \textit{AmountSaved}. Given the right-skewed nature of this measure and that the scale of the amount saved will vary with the baseline valuation, we take the logarithm of the amount saved, and employ it as our dependent variable in an OLS regression. Our coefficients in these regressions thus reflect approximate percentage effects. When performing all estimations, we cluster our standard errors in two dimensions, namely by property account number and by appraiser. 

Subsequently, we turn to the audio recordings. We begin by reporting a visual, descriptive analysis of the measures obtained from the M-LLM, to explore differences in coded appellant behaviors and tones by gender, both of the appellant and the ARB panelist. We then examine how each behavior or tone associates with hearing outcomes, conditional on the appellants' and ARB panelist genders. Finally, we incorporate the behavioral and tonal measures into our baseline regression specification, to understand whether any effects of ARB panelist gender persist conditional on appellant behavior. 

When conducting these final regressions leveraging audio data, we incorporate the appellant gender values that Gemini infers based on the hearing audio. This is important because, as noted above, when dealing with structured hearing records, we gender code only the first name listed among all appellant attendees, and it is possible the coded individual is not the individual who led the conversation. Further, our parsing may be subject to error, given the name format varies across records. And, the gender predictions based on Social Security Administration data may also be occasionally incorrect. Accordingly, by incorporating Gemini's predictions about appellant gender based on voice, we obtain an independent indicator of appellant gender,\footnote{We are somewhat less concerned with gender predictions associated with the ARB panelists because i) formatting of ARB panelist names is consistent throughout the data, suggesting that the values are drawn from an IT system, and ii) the confidence associated with SSA-based gender predictions for ARB panelists is generally high. Further, as mentioned above, we are less comfortable relying on Gemini's gender predictions for ARB panelists, the chair or otherwise, because there are typically three panelists, hence accurate prediction requires both accurate speaker disambiguation/identification followed by a gender inference, implying much more measurement error.} one for which the measurement error is presumably independent of that in our name- and regular-expression-based gender predictions. Most importantly, we can consider estimates from a sub-sample of records among which our name-based and audio-based gender predictions agree, i.e., a sample likely characterized by relatively little measurement error.

Our goals with this second set of regressions, incorporating the data from the audio recordings, are three-fold. First, we hope to ensure that our baseline findings are not a product of endogeneity and bias due to measurement error in appellants' gender predictions obtained via name-parsing using regular expressions. Second, we hope to assess how homeowner behaviors and tone translate to hearing outcomes depending on panel and homeowner gender. Third, and most importantly, we hope to determine whether any systematic differences in hearing outcomes that depend on panel and appellant gender persist after we condition on measures of homeowner behavior and tone during the hearing. Were we to observe the latter, it would suggest that the sources of the gender concordance effects are primarily manifesting via unvoiced factors that are unrelated to appellant behavior (e.g., differences in perceptions or beliefs on the part of ARB panelists).

\vspace{-1mm}

\section{Descriptive Evidence}
\label{sec:desc_results}

Assuming the assignment of ARB panelists is a true experiment, we can perform a simple OLS regression, relating the \textit{AnySaved} dummy to a binary indicator of whether the ARB panel chair is female, absent any controls, and the result will bear a plausibly causal interpretation. Doing so, we obtain the results in Table~\ref{tab:true_exp}, where we observe evidence of a negative gender concordance effect for female appellants, yet no evidence for any such effect among male appellants.\footnote{We have also evaluated these estimates employing randomization inference \citep{young2019channeling}, to ensure robustness. Permuting the \textit{Female Panel Chair} variable 10,000 times across our entire estimation sample and repeating our pair of estimations (one for male appellants and another for female appellants) with each pass, we obtain empirical p-values that indicate a highly significant effect in the case of female appellants (\textit{p} < 0.001), yet statistically insignificant effects in the case of male appellants (\textit{p} > 0.10). Further, we obtain an empirical p-value for the difference between the two estimates that also indicates statistical significance (\textit{p} = 0.0238).}  

\begin{table}[htpb!]
\centering
\caption{Effect of Female Panel Chair on Probability of Any Reduction in Appraisal Value, by Appellant Gender}
\begin{threeparttable}
\label{tab:true_exp}
\begin{tabular}{lcc}
\hline
Appellant Gender & \textbf{Female} & \textbf{Male} \\
\hline
Intercept & 0.653*** (0.006) &  0.656*** (0.006)  \\
\rowcolor{Gray}
Female Panel Chair & -0.016*** (0.006) & -0.004 (0.006)   \\
\hline
Num.Obs. & 39,309 & 75,206 \\
%R2 Adj & 2.435e-4 & 2.46e-6 \\
RMSE & 0.4787 & 0.4757 \\
\bottomrule
\end{tabular}
\begin{tablenotes}
\small
\item Notes: *** p$<$0.01, ** p$<$0.05, * p$<$0.1; Robust standard errors clustered by account and appraiser.  
\end{tablenotes}
\end{threeparttable}
\end{table}

Having established a baseline estimate, we next consider various descriptive analyses that speak to the presence of panelist-specific bias, and whether any such bias is gender dependent. We begin by plotting estimates of the average rate at which each ARB panelist grants home value reductions to appellants, with 95\% confidence intervals. When calculating these values, we limit our attention to ARB panelists who participated in at least 50 hearings during our observation period, to avoid noisy estimates. Further, when calculating these values, we first demean the indicator of \textit{AnySavings} with respect to the tax year, to account for the fact that the rate at which savings were awarded differed systematically over time due to rapid increases in average home value. A cap is imposed on the percentage that property appraisal values can increase from one year to the next. Because home values rose rapidly at some point, property appraisals began to systematically undervalue homes, leading to a decline in the rate at which hearings would result in additional tax savings. Were we to ignore tax year when constructing these estimates, ARB panelists who served in different years would exhibit marked variation in their rate of awarding savings due to these market-level shifts. 

\begin{sidewaysfigure}
\centering
\includegraphics[width=\linewidth]{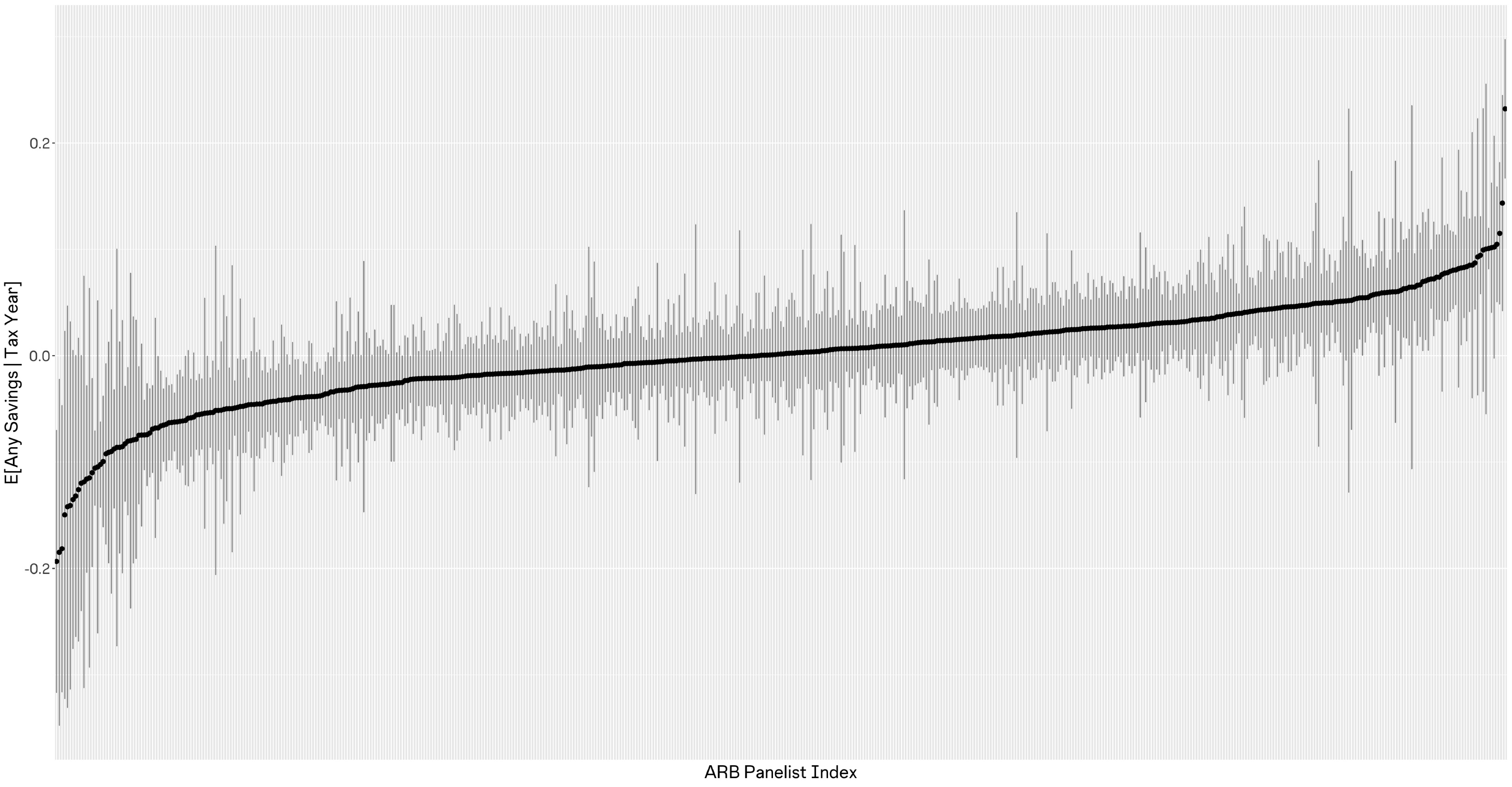}
\caption{Average rate at which tax savings are awarded per ARB panelist, de-meaned by tax year.}
\label{fig:panelist_FEs}
\end{sidewaysfigure}

We depict the resulting estimates in Figure~\ref{fig:panelist_FEs}, ordering panelists along the horizontal axis from least to most likely to adjudicate savings. The average probability that savings are adjudicated spans a range of approximately 40 percentage points, from the most `critical' to the most `favorable' ARB panelist. 

Even if ARB panelists exhibited no systematic bias in their decisions, a normal distribution of savings adjudication rates across panelists might nonetheless be expected due to random variation in case characteristics. Accordingly, to explore whether the observed distribution truly reflects panelist-specific biases in decision-making (i.e., whether the distribution truly differs from what we might expect due to chance), we perform a permutation test \citep{good2013permutation}. We implement our test via a repeated shuffling of the hearing outcome variable (i.e., \textit{AnySaved}) over hearings to recover artificial, simulated ARB panelist-specific estimates of savings adjudication rates under the null hypothesis that adjudication rates are, in fact, homogeneous across panelists. That is, by breaking the relationship between the outcome and assigned ARB panelist identities, we effectively simulate the null distribution of our panelist-specific test statistics distributing the panelist-specific bias uniformly across across panelists. This approach effectively retains the structure of panelist case assignments while enabling us to test the null hypothesis that panelists have no systematic effect on outcomes. 

We summarize the observed degree of panelist-specific bias using the variance of resulting savings adjudication rates, across panelists. It is important to note that our approach provides a conservative test of systematic bias. Variance in the outcome attributable to panelist bias (panelist-specific treatment effects, in a sense) is unaffected by the shuffling procedure; that is, the shuffling procedure does not eliminate the bias, it simply redistributes it across hearings uniformly. Because the variation in the outcome due to bias remains, the distribution we observe under the null reflects an inflated variance relative to what would occur in the true absence of panelist bias. As a result, our permutation test is relatively conservative or robust. In any case, we expect greater variation in savings adjudication rates in the true data, compared to the null distribution, to the extent that panelist-specific bias is at play. 

\begin{sidewaysfigure}
\centering
\includegraphics[width=\linewidth]{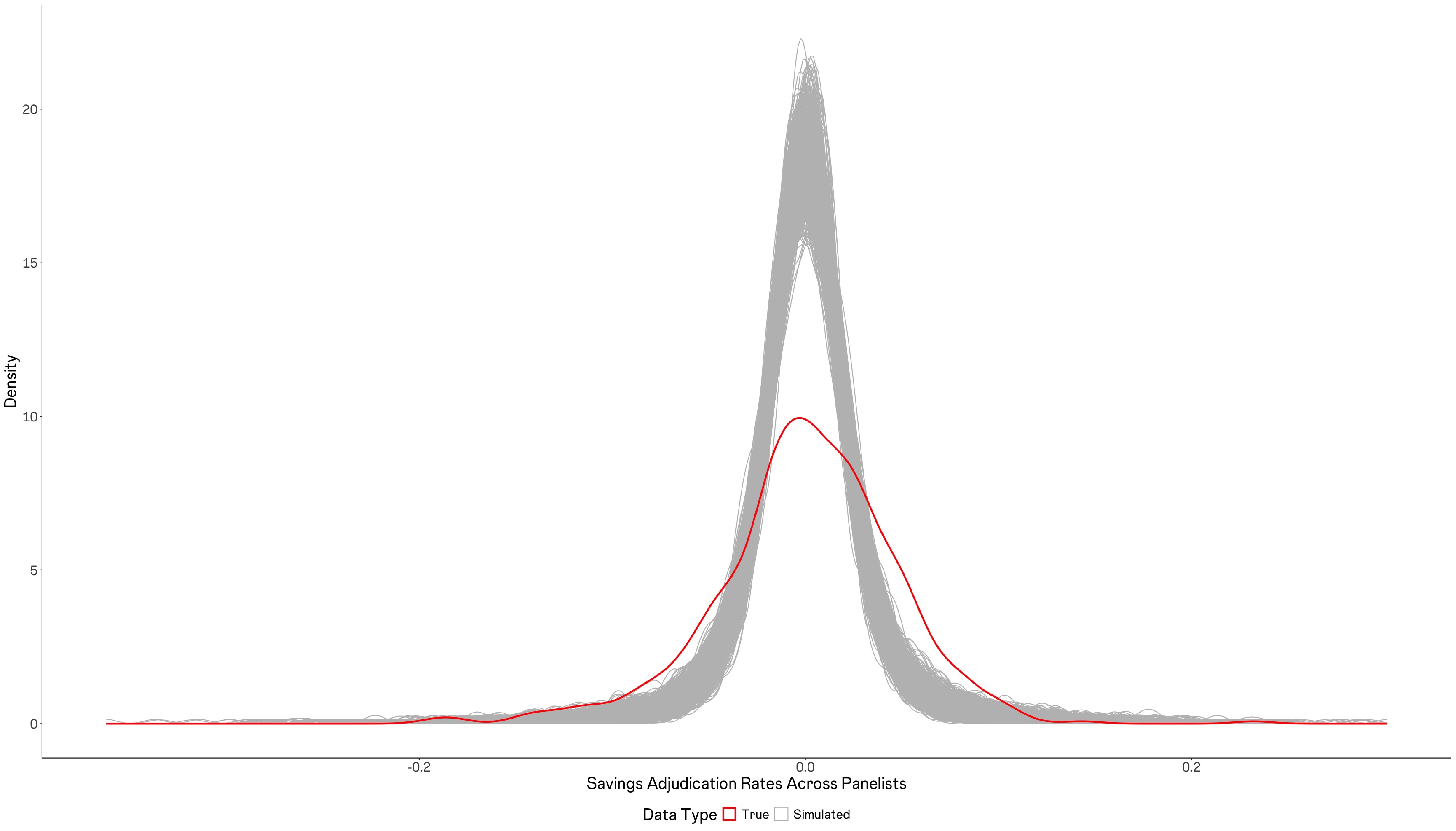}
\caption{Permutation test evaluating panelist bias.}
\label{fig:rand_int}
\end{sidewaysfigure}

In Figure~\ref{fig:rand_int}, we plot the true set of estimates in red, atop the simulated estimates from 1,000 randomly shuffled samples shown in gray. The simulated distributions show what would result from the objective cases -- i.e., the distribution of savings if ARB panelists did not exhibit systematic bias. It is readily apparent that the true distribution has fatter tails than any simulated distribution, such that the dispersion in panelists' rates of savings adjudication is wider than can be explained by chance alone. 

\begin{sidewaysfigure}
\centering
\includegraphics[width=\textwidth]{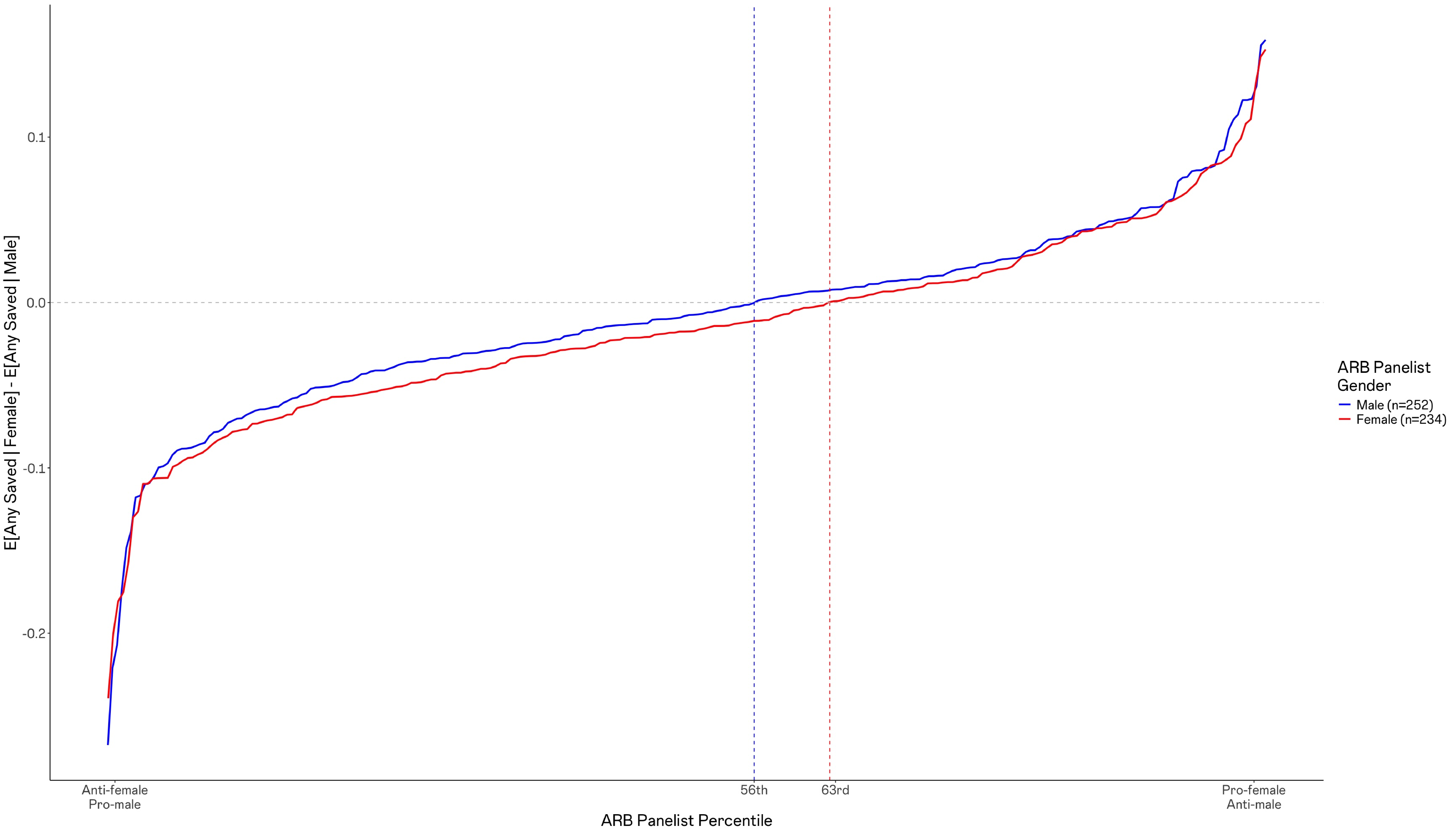}
\caption{\centering Female vs. Male Differential per ARB Panelist}
\label{fig:descriptive_plot}
\end{sidewaysfigure}

Delving deeper into this analysis, we can contrast hearing outcomes per panelist among hearings split by appellant gender, to gain a sense of the extent to which panelists adjudicated decisions might depend on homeowners' gender. We identified all ARB panelists who participated in at least 50 hearings and had also participated in at least 25 hearings involving homeowners of each gender. We calculate the rate at which each panelist awarded a home value reduction, i.e., the mean of our \textit{AnySaved} measure across hearings in which they participated. Finally, we calculate the difference between the resulting values across homeowner genders, i.e., E[AnySaved | Female] - E[AnySaved | Male]. Finally, grouping ARB panelists based on their gender and ordering them from anti-female bias (left) to pro-female bias (right), we plot the resulting value curves in Figure~\ref{fig:descriptive_plot}. 

Inspecting the figure, two facts are immediately apparent. First, a) 55 percent of male panelists exhibit anti-female decision-making, as compared to 62 percent of female panelists, and b) the anti-female tendencies of the most anti-female panelists are much stronger than the anti-male bias of the most anti-male panelists. This latter observation is based on the fact that the most anti-female panelist, on the extreme left of the horizontal axis, adjudicates savings for female appellants with 20\% lower probability than male appellants and, by contrast, the most anti-male panelist, on the extreme right, adjudicates savings for male appellants with approximately 15\% lower probability than female appellants. 

\section{Main Regression Results}
\label{sec:reg_results}

We now turn to our regressions, wherein we can account for changes over time via tax year dummies, and we can also explore within property variation in hearing outcomes across hearings, depending on panel composition. We begin with an estimation of the relationship between ARB board chair gender and hearing outcome, splitting the sample by appellant (homeowner) gender, considering our binary indicator of whether any reduction resulted in assessed home value, i.e., \textit{AnySaved}, employing Ordinary Least Squares (OLS) regression, i.e., a Linear Probability Model (LPM). The results are presented in Table~\ref{tab:baseline_any}. 

\begin{table}[htpb!]
\small
\centering
\caption{Effect of ARB and Appellant Gender on Any Reduction in Home Value}
\label{tab:baseline_any}
\begin{threeparttable}
\begin{tabular}{lcccc}
\toprule
Appellant Gender & \textbf{Female} &\textbf{ Male} & \textbf{Female} & \textbf{Male} \\
\midrule
  \multicolumn{5}{c}{\textbf{Tax Year, Home Value Bucket \& Appraiser FEs}} \\
\midrule
\rowcolor{Gray}
Female Panel Chair  & -0.014*** (0.005) & -0.004 (0.004) &  &  \\
\rowcolor{Gray}
1 Female Panelist &  &  & -0.025*** (0.009) & 0.002 (0.007) \\
\rowcolor{Gray}
2 Female Panelists &  &  & -0.031*** (0.009) & 0.004 (0.008) \\
\rowcolor{Gray}
3 Female Panelists &  &  & -0.042*** (0.010) & -0.007 (0.009) \\
Online & -0.040*** (0.011) & -0.031*** (0.008) & -0.030*** (0.011) & -0.021*** (0.009) \\
Market Equity & 0.115*** (0.014) & 0.125*** (0.009) & 0.116*** (0.014) & 0.127*** (0.009) \\
Unequal Appraisal & 0.036*** (0.006) & 0.035*** (0.005) & 0.037*** (0.006) & 0.036*** (0.005) \\
Count Prior Protests & -0.057*** (0.003) & -0.049*** (0.003) & -0.059*** (0.004) & -0.051*** (0.003) \\
Recent Savings & 0.089*** (0.008) & 0.060*** (0.006) & 0.087*** (0.008) & 0.061*** (0.006) \\
\midrule
Num.Obs. & 39,309 & 75,206 & 35,453 & 67,732 \\
R2 & 0.070 & 0.070 & 0.063 & 0.059 \\
R2 Adj. & 0.061 & 0.065 & 0.053 & 0.054 \\
RMSE & 0.46 & 0.46 & 0.46 & 0.45 \\
\midrule
 \multicolumn{5}{c}{\textbf{Tax Year, Account \& Appraiser FEs}} \\
\midrule
\rowcolor{Gray}
Female Panel Chair & -0.020** (0.010) & -0.005 (0.006) &  &  \\
\rowcolor{Gray}
1 Female Panelist &  &  & -0.044*** (0.017) & 0.015 (0.011) \\
\rowcolor{Gray}
2 Female Panelists &  &  & -0.036*** (0.017) & 0.018 (0.011) \\
\rowcolor{Gray}
3 Female Panelists &  &  & -0.048*** (0.020) & 0.010 (0.013) \\
Online & -0.007 (0.022) & 0.014 (0.015) & 0.006 (0.024) & 0.024 (0.016) \\
Market Equity & 0.066*** (0.022) & 0.061*** (0.016) & 0.062*** (0.023) & 0.061*** (0.016) \\
Unequal Appraisal & 0.020 (0.013) & 0.045*** (0.008) & 0.014 (0.014) & 0.049*** (0.008) \\
Count Prior Protests & -0.037*** (0.009) & -0.020*** (0.005) & -0.037*** (0.009) & -0.019*** (0.006) \\
Recent Savings & -0.239*** (0.010) & -0.268*** (0.006) & -0.243*** (0.010) & -0.268*** (0.007) \\
\midrule
Num.Obs. & 39,309 & 75,206 & 35,453 & 67,732 \\
R2 & 0.862 & 0.829 & 0.863 & 0.829 \\
R2 Adj. & 0.270 & 0.285 & 0.260 & 0.266 \\
RMSE & 0.18 & 0.20 & 0.18 & 0.19 \\
\midrule
Mean(Any Saved) & 0.644  & 0.654 & 0.661 & 0.674 \\
\bottomrule
\end{tabular}
\begin{tablenotes}
\small
\item Notes: *** p$<$0.01, ** p$<$0.05, * p$<$0.1; Robust standard errors clustered by account and appraiser; Observation counts decline in columns 3 and 4 due to the omission of hearings involving only one ARB member.  
\end{tablenotes}
\end{threeparttable}
\end{table}

The upper panel presents regression results from models that include fixed effects by tax year, home value buckets, and appraiser, whereas the lower panel replaces home value bucket fixed effects by account fixed effects.\footnote{We were unable to obtain a response from the Harris County Appraisal District, but speaking with representatives of the Dallas County Appraisal District, they indicated that the property account number remains tied to a particular home even if it changes owners. The DCAD representative indicated they believed the same was true of all counties in Texas.} We observe consistent results across the board in terms of gender effects. The probability that female homeowners obtain reductions in their home value assessment, and thus their property taxes, is systematically lower (higher) when they face a female (male) ARB chair. Considering the magnitude, the difference in probability is approximately 1.4 pp. Considering three-member panel hearings only, and replacing the binary indicator of ARB chair gender with dummies reflecting the count of female panelists (the reference value is 0, implying an all-male panel), we see the same pattern. Female homeowners' probability of obtaining any tax savings is monotonically decreasing in the number of female panelists they face, such that the difference in the probability of tax savings between panels comprised entirely of males versus entirely of females is approximately 4.2 pp. Next, we consider the bottom panel of estimates, where we incorporate home account fixed effects, enabling us to focus strictly on variation in the panel gender composition across multiple hearings that relate to the same home. We again observe consistent results, though statistical significance declines, perhaps unsurprisingly. Importantly, by contrast, when we consider gender concordance effects for male homeowners, we observe no statistically significant effects, in any of the estimations. Indeed, all coefficients associated with our chair and panelist gender dummies indicate precisely estimated null results. 

Beyond the gender effects, our other coefficients reveal interesting patterns. First, online hearings are systematically less likely to result in tax savings, though these effects disappear following our inclusion of account fixed effects, which suggests that those results may be driven by self-selection. For example, perhaps homeowners are more likely to request a hearing online when they anticipate their case is weaker and less likely to result in savings. 

Second, we observe that, when appellants have previously protested and obtained any savings in their most recent hearing, this generally associates positively with current hearing outcomes when we do not include an account fixed effect. Once we introduce the account fixed effect, the coefficient turns negative. These results suggest that protest hearings involving certain owners, appellants, or property characteristics are systematically more likely to achieve reductions to an initial appraisal value. However, once we account for those stable, cross-sectional differences, recent success in a protest hearing reduces the likelihood of subsequent favorable outcomes. That is, a fresh win appears to reduce the likelihood of obtaining another reduction. This dynamic can likely be explained by the fact that recent savings achieved in a prior protest will generally imply that some error has been identified in the prior appraisal process, and that error has been rectified. Once errors have been addressed, the chance of success in subsequent protests will naturally decline, as arguments for any reduction will be reduced. Third, we observe a consistent negative effect of prior protest hearing counts in all regressions. Given our estimations condition on whether a property has obtained savings on the most recent prior appeal, this effect could indicate that appraisal review board panels, observing a lengthy history of prior protests within the property's file, may view the appellant's case with greater skepticism.

That said, it is important to also recognize that several administrative restrictions limit how property taxes can be adjusted. These restrictions influence the incentives for households to appeal an appraisal, as well as the potential savings households may obtain from an appeal. Further, these restrictions make it difficult to evaluate the temporal evolution of appeals and savings arising from them \citep{giaccobasso2022my}. Most notably, property tax increases are capped at 10\% annually. Households already under the cap have less incentive to appeal since their efforts would not immediately reduce their property taxes. 

We next explore our measure of the dollar amount saved. As noted earlier, these estimations employ the logarithm of the amount saved as the outcome. The coefficient estimates are thus interpretable as approximate percentage effects. Considering Table~\ref{tab:baseline_amt}, we observe the same general pattern of effects as in Table~\ref{tab:baseline_any}. Whereas female homeowners' case outcomes differ systematically depending on the gender of the ARB panelists, outcomes for male homeowners do not. Considering our estimations in column 1, which employs the same right-hand side model as in column 1 of Table~\ref{tab:baseline_any}, we see that female homeowners save approximately 12.2\% less (more) when faced with a female (male) ARB panel chair.\footnote{exp(-0.13) - 1 = -0.122} Further, we see that the amount saved declines with the addition of more female panelists, such that the difference in home value reduction obtained by female homeowners when facing an all-female versus all-male panel is approximately 32.8\%.\footnote{exp(-0.397) - 1 = -0.328} Incorporating account fixed effects, we observe consistent findings in the case of ARB chair gender, though statistical significance declines slightly. 

\begin{table}[htpb!]
\small
\centering
\caption{Effect of ARB and Appellant Gender on Log(Dollar Reduction in Home Value)}
\label{tab:baseline_amt}
\begin{threeparttable}
\begin{tabular}{lcccc}
\toprule
Appellant Gender & \textbf{Female} & \textbf{Male} & \textbf{Female} & \textbf{Male} \\
\midrule
\multicolumn{5}{c}{\textbf{Tax Year, Home Value Bucket \& Appraiser FEs}} \\
\midrule
\rowcolor{Gray}
Female Panel Chair & -0.131*** (0.053) & -0.062 (0.045) &  &  \\
\rowcolor{Gray}
Count Female Panelists: 1 &  &  & -0.233*** (0.092) & 0.004 (0.073) \\
\rowcolor{Gray}
Count Female Panelists: 2 &  &  & -0.304*** (0.087) & 0.009 (0.078) \\
\rowcolor{Gray}
Count Female Panelists: 3 &  &  & -0.397*** (0.102) & -0.103 (0.087) \\
Online & -0.436*** (0.101) & -0.447*** (0.080) & -0.363*** (0.110) & -0.359*** (0.085) \\
Market Equity & 1.353*** (0.139) & 1.549*** (0.094) & 1.361*** (0.138) & 1.561*** (0.094) \\
Unequal Appraisal & 0.203*** (0.055) & 0.215*** (0.052) & 0.206*** (0.056) & 0.219*** (0.052) \\
Count Prior Protests & -0.553*** (0.033) & -0.491*** (0.027) & -0.568*** (0.034) & -0.512*** (0.027) \\
Recent Savings & 0.918*** (0.075) & 0.632*** (0.056) & 0.894*** (0.079) & 0.640*** (0.058) \\
\midrule
Num.Obs. & 39,309 & 75,206 & 35,453 & 67,732 \\
R2 & 0.082 & 0.084 & 0.077 & 0.075 \\
R2 Adj. & 0.073 & 0.079 & 0.067 & 0.070 \\
RMSE & 4.50 & 4.53 & 4.46 & 4.49 \\
\midrule
\multicolumn{5}{c}{\textbf{Tax Year, Account \& Appraiser FEs}} \\
\midrule
\rowcolor{Gray}
Female Panel Chair & -0.211*** (0.098) & -0.104 (0.064) &  &  \\
\rowcolor{Gray}
Count Female Panelists: 1 &  &  & -0.400*** (0.164) & 0.066 (0.115) \\
\rowcolor{Gray}
Count Female Panelists: 2 &  &  & -0.334*** (0.168) & 0.094 (0.108) \\
\rowcolor{Gray}
Count Female Panelists: 3 &  &  & -0.476*** (0.194) & 0.006 (0.128) \\
Online & -0.071 (0.214) & 0.055 (0.149) & 0.086 (0.229) & 0.154 (0.167) \\
Market Equity & 0.753*** (0.219) & 0.864*** (0.153) & 0.698*** (0.232) & 0.850*** (0.159) \\
Unequal Appraisal & 0.158 (0.131) & 0.430*** (0.079) & 0.080 (0.137) & 0.455*** (0.084) \\
Count Prior Protests & -0.421*** (0.089) & -0.252*** (0.053) & -0.406*** (0.093) & -0.246*** (0.060) \\
Recent Savings & -2.320*** (0.100) & -2.662*** (0.065) & -2.390*** (0.104) & -2.670*** (0.068) \\
\midrule
Num.Obs. & 39,309 & 75,206 & 35,453 & 67,732 \\
R2 & 0.862 & 0.826 & 0.863 & 0.826 \\
R2 Adj. & 0.271 & 0.270 & 0.261 & 0.253 \\
RMSE & 1.74 & 1.98 & 1.72 & 1.95 \\
\midrule
Mean(Amount Saved) & \$20,206.00  & \$24,379.00 & \$20,678.00  & \$24,904.00  \\
\bottomrule
\end{tabular}
\begin{tablenotes}
\small
\item Notes: *** p$<$0.01, ** p$<$0.05, * p$<$0.1; Robust standard errors clustered by account and appraiser; Observation counts decline in columns 3 and 4 due to the omission of hearings involving only one ARB member. Log transformation is applied as Ln(y+1) in the case of \$0 savings. 
\end{tablenotes}
\end{threeparttable}
\end{table}

\section{Audio Analysis}
\label{sec:audio_analysis}

We next leverage a multi-modal large language model (LLM), specifically Gemini 1.5 Flash, to process all audio recordings associated with nearly all owner-attended ARB hearings in our original sample. Ultimately, we work with a subset of 80,197 audio files, namely those we can reliably match to hearings from our original sample of structured hearing administrative records. We are unable to arrive at matches for all audio files and all structured hearing records because some structured hearing records in our sample do not have associated audio recordings. Further, in some cases, we have multiple structured hearing records for the same property in the same tax year (e.g., if separate hearings were held to adjudicate different protest reasons). In such cases, we lack any means of correctly discerning which audio file is associated with which structured hearing record. In all such cases, we omit the relevant audio files and structured hearing records from our analyses of in-hearing behavior. 

Employing Gemini, we construct measures that quantify appellant behavior during a hearing. Using this data, we hope to understand i) any systematic differences that female homeowners may exhibit depending on the gender composition of the panel they face, and ii) any systematic differences in the relationship between homeowners' behavior and hearing outcomes, depending on ARB panelist gender. 

As noted earlier, we consider a measure of hearing length (in seconds), extracted from the metadata of each audio recording. Further, we consider Gemini's assessment of the fraction of time an appellant spent talking during a hearing, binary indicators of how the appellant sounds, e.g., confident, frustrated, nervous, defensive, etc., and binary indicators of appellant actions during the hearing, e.g., referencing evidence, raising their voice, interrupting a panelist, disagreeing with a panelist, asking questions, asking a panelist to speak louder, or asking a panelist to repeat themselves.

\subsection{The Quality of LLMs' Audio Coding}

Before considering the results, it is worth considering the time and cost savings from relying on a multimodal LLM for this annotation exercise. The average recording is 15 minutes long. The total length of audio is approximately 2.8 years in length. Had we hired a human annotator for the task, paying minimum wage, i.e., \$7.25/hr, the total cost would have amounted to approximately \$177,000. In contrast, our total cost of employing Gemini 1.5 Flash for this exercise was approximately \$450. The time to complete this annotation task would also have required years of effort (perhaps weeks or months, in practice, were we to have employed many coders in parallel). By contrast, processing recordings sequentially via the Gemini API (i.e., without parallel processing) required approximately 14 days. With parallel processing, the task could have been completed in a matter of days.

Before we proceed, it is helpful to obtain some sense of Gemini's accuracy. We have a couple of Gemini-coded values that we can compare against `ground truth' labels of a sort, derived from the structured hearing data. In particular, we examine Gemini's inferences about whether the appellant seems to be the property owner versus an agent hired on the owner's behalf, where the reality is that all appellants in our data are property owners. Additionally, we examine Gemini's inferences about whether a hearing sounds like it took place online or in person, comparing the result with the actual medium recorded in the hearing record. The results of these comparisons appear in Table~\ref{tab:virtual_vs_online} and \ref{tab:appellant_property_ownership}. 

Considering Gemini's ability to infer appellants' online attendance at their hearing, it is important to recognize that this inference depends on explicit cues in the hearing audio, e.g., appellant references to their screen sharing and audio configuration. To the extent an appellant attending online has no difficulty with their technology setup, cues of online attendance will be absent. Conversely, for hearings that truly took place offline, there should never be cues of online attendance. Our result here is entirely consistent with that idea. Among hearings that took place offline, it makes sense that no such cues would be present. Accordingly, as we would hope, Gemini rarely infers that an offline hearing occurred online. Gemini mistakenly labels offline hearings as online in just 1.2\% of instances (827 out of 69,312 hearings). By contrast, among online hearings, Gemini fails to detect a verbal cue indicating online attendance by the appellant in approximately 39.8\% of instances (4,334 out of 10,885 hearings).

\begin{table}[htpb!]
\centering
\caption{Cross-tab of Actual vs. Inferred Online vs. Offline Hearing Attendance}
\begin{tabular}{lcc}
\toprule
 & Gemini Offline & Gemini Online \\
\midrule
True Offline & 68,485 & 827 \\
True Online & 4,334 & 6,551 \\
\bottomrule
\end{tabular}
\label{tab:virtual_vs_online}
\end{table}

Considering Gemini's inferences about whether the appellant is the property owner, again, this will depend on explicit reference being made to the appellant's ownership of the property, or some other indicative statements, serving as cues that the appellant lives at the property. Such cues should be relatively common, as ARB panels often ask the attendees whether they own the property. Further, appellants will often reference living at the location, the price at which they purchased the property, and so on. It is thus reassuring to see that Gemini correctly identifies the appellant as the property owner in approximately 92.8\% of our hearings sample (as noted earlier, our sample is exclusively comprised of owner-attended hearings).

\begin{table}[htpb!]
\centering
\caption{Gemini-inferred Property Owner vs. Hired Agent}
\begin{tabular}{lc}
\toprule
Gemini's Inference & Count \\
\midrule
Agent & 5,758 \\ 
Owner & 74,439 \\
\bottomrule
\end{tabular}
\label{tab:appellant_property_ownership}
\end{table}

Note that we also obtained annotations from human coders for a random sample of recordings, for comparison. That analysis, reported in Appendix B, suggests that Gemini's annotations do contain meaningful information.

\subsection{Descriptive Results}

\begin{figure}[htpb!]
\centering
\includegraphics[width=1.1\textwidth]{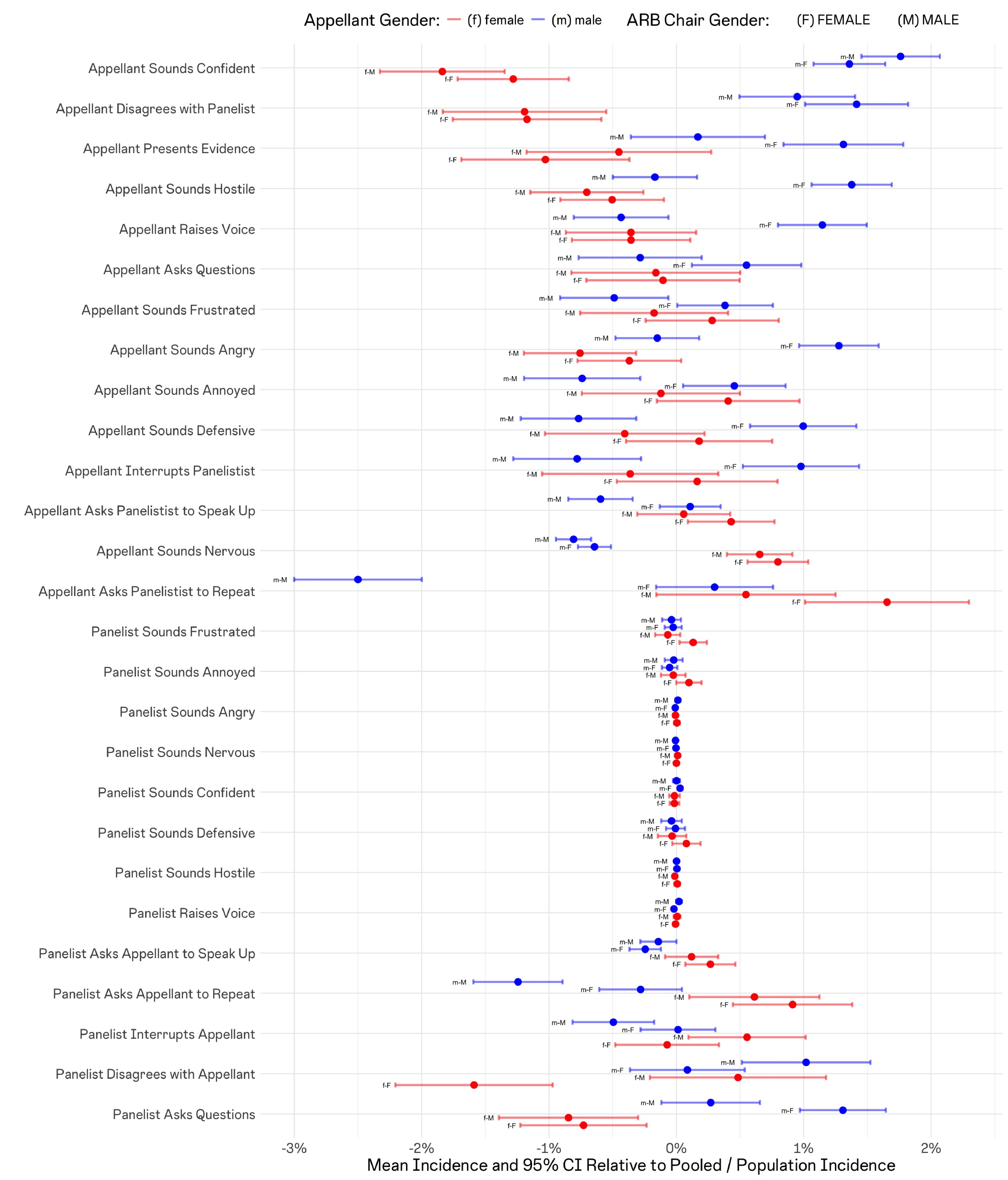}
\caption{\centering Appellant and Panelist Behavior by Own and ARB Panel Chair Gender. Panelist Behavior Refers to the Panelist Who 'Speaks the Most' During the Hearing, and Appellant Behavior Refers to the Appellant Who 'Speaks the Most' During the Hearing. We explain the definition of each tone and behavior in the Appendix.}
\label{fig:gemini_behavior}
\end{figure}

We now turn to Gemini's responses about the appellants' and panelists' behaviors and tone. In Figure~\ref{fig:gemini_behavior} we graphically depict the incidence rate for each behavior or tone for appellants and panelists, reporting split sample means and 95\% confidence intervals for each combination of attendee gender and ARB chair gender. The values we depict are re-centered around the population mean incidence for the respective behavior or tone, to facilitate easier comparisons of cross-gender differences. For example, we observe that, in the case of male appellants facing male ARB chairs, appellants are systematically more likely to sound confident (they are labeled as such 1.75\% more often than the average incidence across our entire sample). In contrast, female appellants are systematically less likely to sound confident when facing a male ARB chair. This annotation occurs nearly 2\% less often than the overall incidence in the population. 

Several notable patterns are immediately apparent from the plot. First, as we observed in our descriptive statistics for the labels, the coded behavior and tones of ARB panelists exhibit much less variance than those of appellants. Across most measures of behavior and tone, Gemini codes panelists quite homogeneously across hearings. The only exceptions relate to our measures of whether panelists ask the appellant to 'Speak Up', ask them to 'Repeat' themselves, 'Interrupt' them, or disagree with them. Of greatest note is that female panelists are systematically less likely to verbally disagree with a female appellant, which contradicts what we might expect given that female appellants tend to fare worse with female ARB panels.   

Considering the appellants' behavior, we observe immediate and stark differences between males and females. For example, male appellants are less likely to be coded as sounding nervous and more likely to be coded as sounding confident. Males are also more likely than females to verbally disagree with the ARB panel, they are also more likely to present formal evidence, and they are more likely to raise their voice or sound hostile. Interestingly, however, some of these more aggressive behaviors are mainly targeted toward female ARB panelists. Considering how appellants' behavior \textit{differs} depending on ARB chair gender, we see that male appellants are more likely to interrupt female panelists, more likely to raise their voice toward female panelists, more likely to sound hostile toward female panelists, and more likely to sound defensive, compared to their behavior when facing a male ARB panelist.

By contrast, female appellants exhibit relatively few differences in their behavior between male and female ARB panelists. Even for those behaviors and tones where female appellants' behavior does differ based on ARB chair gender, the differences are relatively small in magnitude, e.g., no more than a 0.5 pp difference in the probability a behavior or tone is exhibited. Further, the differences are statistically significant in only four cases: i) asking a panelist to repeat themselves, ii) asking a panelist to 'speak up', and iii) sounding confident. Notably, however, in these cases, the behavior or tone in question is more pronounced when a female appellant faces a female rather than a male ARB chair. As such, we do observe evidence of some small behavioral and tonal differences on the part of female appellants depending on the ARB panel chair's gender that \textit{may} explain the less favorable outcomes of female appellants, to some extent, though the relatively small degree of variation in behavior suggests that this explanation is unlikely to explain differential hearing outcomes entirely. 

Before we proceed further, it is also helpful to consider the contrast between our regular-expression, Social Security Administration-based predictions of appellant genders and the predictions that Gemini makes based on audible voice in the recordings. As noted earlier, we would anticipate that measurement error will arise with each approach, but that the errors between each measure should be independent of one another. Whereas our regular-expression approach focused on the first name appearing in the formal attendee field, our request of Gemini is to predict the gender of the appellant who 'speaks the most' during the hearing. The order of names entered in the structured data is not necessarily reflective of which appellant led the case presentation at the hearing. That said, for Gemini to succeed at this task, it is not only required to discern gender based on voice, it must also correctly identify which speaker is the appellant. Moreover, even the ultimate task of inferring gender from the vocal pitch, etc., for the identified speaker can be challenging. The audio recordings vary significantly in their quality, and for some voices, even with clear audio, it may be difficult to determine a speaker's gender with any degree of confidence \citep[see ][for a lengthy discussion of gender inference around audible vocal characteristics]{schanke2024digital}. We present the cross-tabulation of Gemini and regular-expression-based gender inferences for appellants in Table~\ref{tab:gender_pred_crosstab}, where we see that the two approaches agree with one another in 77.8\% of cases. This high level of agreement suggests that there is more signal than noise in each approach we take, yet, as expected, there also appears to be substantial error with each. 

\begin{table}[htpb!]
\centering
\caption{Cross-tabulation of Gender Inference: Gemini (Audio) vs. Regular Expression (SSA)}
\begin{tabular}{lcc}
\toprule
 & Regular Expression: Female & Regular Expression: Male \\
\midrule
Gemini: Female & 17,077 & 7,145 \\
Gemini: Male   & 10,628 & 45,347 \\
\bottomrule
\end{tabular}
\label{tab:gender_pred_crosstab}
\end{table}

Given the above, before we proceed with our analysis of the impact of appellant behaviors on hearing outcomes, we will first explore the sensitivity of our earlier findings by relying on either measure, separately, or their intersection. We first repeat the baseline regressions on our audio-recorded sample, splitting the sample based on appellant gender as determined by Gemini from audio. The results, presented in Table~\ref{tab:baseline_gemini_any}, are generally consistent with our earlier findings. The estimates are slightly attenuated relative to our earlier estimates, but the general pattern of effects remains consistent. Further, in Table~\ref{tab:baseline_gemini_regex_any}, we again repeat the estimations, this time employing only those hearings where Gemini's prediction about appellant gender agrees with our regular-expression-based predictions. Again, we find consistent results. Repeating the same exercise with our $Amount Saved$ outcome, we obtain the results in Tables~\ref{tab:baseline_gemini_amt} and \ref{tab:baseline_gemini_regex_amt}. Again, we observe consistent results in each case. Broadly, the stability of our estimates is reassuring, as it suggests that measurement error in the gender predictions, whatever its nature, does not drive or materially influence our findings.  

\begin{table}[htpb!]
\small
\centering
\caption{Effect of ARB Gender on Home Value Reduction (DV = AnySaved) by (Voice-based, Gemini-detected) Appellant Gender}
\label{tab:baseline_gemini_any}
\begin{threeparttable}
\begin{tabular}{lcccc}
\toprule
Appellant Gender & \textbf{Female} & \textbf{Male} & \textbf{Female} & \textbf{Male} \\
\midrule
  \multicolumn{5}{c}{\textbf{Tax Year, Home Value Bucket \& Appraiser FEs}} \\
\midrule
\rowcolor{Gray}
Female Panel Chair & -0.011* (0.006) & -0.003 (0.005) &  &  \\
\rowcolor{Gray}
Female Panelists (1) &  &  & -0.016 (0.010) & -0.002 (0.008) \\
\rowcolor{Gray}
Female Panelists (2) &  &  & -0.020* (0.011) & 0.003 (0.007) \\
\rowcolor{Gray}
Female Panelists (3) &  &  & -0.038*** (0.012) & -0.006 (0.008) \\
Online & -0.042*** (0.012) & -0.014 (0.010) & -0.030** (0.014) & -0.005 (0.010) \\
Market Equity & 0.131*** (0.016) & 0.113*** (0.009) & 0.133*** (0.016) & 0.113*** (0.009) \\
Unequal Appraisal & 0.035*** (0.007) & 0.038*** (0.005) & 0.037*** (0.007) & 0.038*** (0.005) \\
Count Prior Protests & -0.056*** (0.004) & -0.054*** (0.003) & -0.060*** (0.005) & -0.057*** (0.003) \\
Saved Last Protest & 0.076*** (0.009) & 0.074*** (0.006) & 0.076*** (0.009) & 0.077*** (0.006) \\
\midrule
Num.Obs. & 24,222 & 55,975 & 22,030 & 52,674 \\
R2 & 0.07 & 0.055 & 0.064 & 0.051 \\
R2 Adj. & 0.056 & 0.048 & 0.049 & 0.044 \\
RMSE & 0.46 & 0.45 & 0.46 & 0.45 \\
\midrule
  \multicolumn{5}{c}{\textbf{Tax Year, Account and Appraiser FEs}} \\
\midrule
\rowcolor{Gray}
Female Panel Chair & -0.028* (0.015) & 0.004 (0.008) &  &  \\
\rowcolor{Gray}
Female Panelists (1) &  &  & -0.027 (0.025) & 0.009 (0.015) \\
\rowcolor{Gray}
Female Panelists (2) &  &  & -0.040 (0.025) & 0.021 (0.014) \\
\rowcolor{Gray}
Female Panelists (3) &  &  & -0.024 (0.030) & 0.013 (0.016) \\
Online & 0.090** (0.036) & 0.018 (0.020) & 0.058 (0.042) & 0.030 (0.023) \\
Market Equity & 0.096*** (0.034) & 0.042* (0.022) & 0.102*** (0.035) & 0.044** (0.022) \\
Unequal Appraisal & 0.052** (0.021) & 0.040*** (0.010) & 0.055** (0.022) & 0.040*** (0.010) \\
Count Prior Protests & -0.016 (0.013) & -0.019*** (0.007) & -0.012 (0.014) & -0.021*** (0.007) \\
Saved Last Protest & -0.240*** (0.016) & -0.268*** (0.009) & -0.240*** (0.017) & -0.270*** (0.009) \\
\midrule
Num.Obs. & 24,222 & 55,975 & 22,030 & 52,674 \\
R2 & 0.91 & 0.863 & 0.91 & 0.862 \\
R2 Adj. & 0.247 & 0.266 & 0.241 & 0.257 \\
RMSE & 0.14 & 0.17 & 0.14 & 0.17 \\
\midrule
Mean(Any Saved) &  0.647 & 0.677 & 0.664 & 0.687 \\
\bottomrule
\end{tabular}
\begin{tablenotes}
\small
\item Notes: *** p$<$0.01, ** p$<$0.05, * p$<$0.1; Robust standard errors clustered by account and appraiser; Observation counts decline in columns 3 and 4 due to the omission of hearings involving only one ARB member. 
\end{tablenotes}
\end{threeparttable}
\end{table}

\begin{table}[htpb!]
\centering
\small
\caption{Effect of ARB on Home Value Reduction (DV = AnySaved) by (Gemini-detected = Regex-based) Appellant Gender}
\begin{tabular}{lcccc}
\toprule
Appellant Gender & \textbf{Female} & \textbf{Male} & \textbf{Female} & \textbf{Male} \\
\midrule
  \multicolumn{5}{c}{\textbf{Tax Year, Home Value Bucket \& Appraiser FEs}} \\
\midrule
\rowcolor{Gray}
Female Panel Chair & -0.016** (0.008) & 0.000 (0.005) &  &  \\
\rowcolor{Gray}
Female Panelists (1) &  &  & -0.027** (0.012) & 0.003 (0.009) \\
\rowcolor{Gray}
Female Panelists (2) &  &  & -0.038*** (0.013) & 0.010 (0.008) \\
\rowcolor{Gray}
Female Panelists (3) &  &  & -0.042*** (0.013) & 0.002 (0.009) \\
Online & -0.053*** (0.015) & -0.010 (0.011) & -0.048*** (0.017) & -0.001 (0.012) \\
Market Equity & 0.127*** (0.019) & 0.116*** (0.010) & 0.129*** (0.019) & 0.117*** (0.010) \\
Unequal Appraisal & 0.035*** (0.008) & 0.039*** (0.005) & 0.037*** (0.008) & 0.039*** (0.005) \\
Count Prior Protests & -0.057*** (0.005) & -0.052*** (0.003) & -0.059*** (0.006) & -0.055*** (0.003) \\
Saved Last Protest & 0.077*** (0.011) & 0.068*** (0.006) & 0.074*** (0.011) & 0.070*** (0.006) \\
\midrule
Num.Obs. & 17,077 & 45,347 & 15,577 & 42,497 \\
R2 & 0.079 & 0.058 & 0.075 & 0.054 \\
R2 Adj. & 0.060 & 0.050 & 0.054 & 0.045 \\
RMSE & 0.46 & 0.45 & 0.45 & 0.45 \\
\midrule
  \multicolumn{5}{c}{\textbf{Tax Year, Account and Appraiser FEs}} \\
\midrule
\rowcolor{Gray}
Female Panel Chair & -0.039** (0.019) & 0.008 (0.009) &  &  \\
\rowcolor{Gray}
Female Panelists (1) &  &  & -0.043 (0.032) & 0.018 (0.016) \\
\rowcolor{Gray}
Female Panelists (2) &  &  & -0.063* (0.032) & 0.029* (0.015) \\
\rowcolor{Gray}
Female Panelists (3) &  &  & -0.052 (0.038) & 0.026 (0.018) \\
Online & 0.102** (0.049) & 0.023 (0.024) & 0.104* (0.055) & 0.028 (0.027) \\
Market Equity & 0.019 (0.047) & 0.043* (0.024) & 0.022 (0.048) & 0.046* (0.024) \\
Unequal Appraisal & 0.025 (0.027) & 0.045*** (0.011) & 0.029 (0.028) & 0.046*** (0.011) \\
Count Prior Protests & -0.039** (0.016) & -0.015* (0.008) & -0.028 (0.018) & -0.013 (0.008) \\
Saved Last Protest & -0.233*** (0.020) & -0.269*** (0.010) & -0.240*** (0.022) & -0.272*** (0.010) \\
\midrule
Num.Obs. & 17,077 & 45,347 & 15,577 & 42,497 \\
R2 & 0.923 & 0.867 & 0.924 & 0.866 \\
R2 Adj. & 0.226 & 0.274 & 0.224 & 0.264 \\
RMSE & 0.13 & 0.17 & 0.13 & 0.17 \\
\midrule
Mean(Any Saved) & 0.650 & 0.679 & 0.665 & 0.689 \\
\bottomrule
\end{tabular}
\begin{tablenotes}
\small
\item Notes: *** p$<$0.01, ** p$<$0.05, * p$<$0.1; Robust standard errors clustered by account and appraiser; Observation counts decline in columns 3 and 4 due to the omission of hearings involving only one ARB member. 
\end{tablenotes}
\label{tab:baseline_gemini_regex_any}
\end{table}

\begin{table}[htpb!]
\small
\centering
\caption{Effect of ARB Gender on Home Value Reduction (DV = Log(AmtSaved+1) by (Voice-based, Gemini-detected) Appellant Gender}
\label{tab:baseline_gemini_amt}
\begin{threeparttable}
\begin{tabular}{lcccc}
\toprule
Appellant Gender & \textbf{Female} & \textbf{Male} & \textbf{Female} & \textbf{Male} \\
\midrule
  \multicolumn{5}{c}{\textbf{Tax Year, Home Value Bucket \& Appraiser FEs}} \\
\midrule
\rowcolor{Gray}
Female Panel Chair & -0.140** (0.063) & -0.037 (0.046) &  &  \\
\rowcolor{Gray}
Female Panelists (1) &  &  & -0.155 (0.099) & -0.021 (0.078) \\
\rowcolor{Gray}
Female Panelists (2) &  &  & -0.233** (0.105) & 0.022 (0.077) \\
\rowcolor{Gray}
Female Panelists (3) &  &  & -0.399*** (0.121) & -0.075 (0.081) \\
Online & -0.438*** (0.117) & -0.263*** (0.097) & -0.325** (0.139) & -0.177* (0.104) \\
Market Equity & 1.484*** (0.158) & 1.394*** (0.091) & 1.496*** (0.158) & 1.398*** (0.092) \\
Unequal Appraisal & 0.221*** (0.065) & 0.216*** (0.047) & 0.232*** (0.067) & 0.214*** (0.047) \\
Count Prior Protests & -0.558*** (0.041) & -0.544*** (0.030) & -0.595*** (0.045) & -0.573*** (0.030) \\
Saved Last Protest & 0.770*** (0.082) & 0.777*** (0.059) & 0.758*** (0.086) & 0.799*** (0.058) \\
\midrule
Num.Obs. & 24,222 & 55,975 & 22,030 & 52,674 \\
R2 & 0.083 & 0.067 & 0.078 & 0.064 \\
R2 Adj. & 0.069 & 0.060 & 0.063 & 0.057 \\
RMSE & 4.47 & 4.48 & 4.43 & 4.45 \\
\midrule
  \multicolumn{5}{c}{\textbf{Tax Year, Account and Appraiser FEs}} \\
\midrule
\rowcolor{Gray}
Female Panel Chair & -0.326** (0.151) & -0.028 (0.082) &  &  \\
\rowcolor{Gray}
Female Panelists (1) &  &  & -0.235 (0.238) & 0.035 (0.146) \\
\rowcolor{Gray}
Female Panelists (2) &  &  & -0.395 (0.245) & 0.110 (0.139) \\
\rowcolor{Gray}
Female Panelists (3) &  &  & -0.279 (0.284) & 0.021 (0.165) \\
Online & 0.844** (0.355) & 0.155 (0.200) & 0.586 (0.410) & 0.211 (0.234) \\
Market Equity & 1.100*** (0.331) & 0.637*** (0.215) & 1.155*** (0.339) & 0.665*** (0.219) \\
Unequal Appraisal & 0.487** (0.206) & 0.367*** (0.102) & 0.495** (0.217) & 0.356*** (0.101) \\
Count Prior Protests & -0.201 (0.127) & -0.236*** (0.069) & -0.156 (0.138) & -0.236*** (0.077) \\
Saved Last Protest & -2.354*** (0.153) & -2.627*** (0.085) & -2.363*** (0.164) & -2.662*** (0.088) \\
\midrule
Num.Obs. & 24,222 & 55,975 & 22,030 & 52,674 \\
R2 & 0.909 & 0.861 & 0.909 & 0.860 \\
R2 Adj. & 0.244 & 0.254 & 0.235 & 0.245 \\
RMSE & 1.41 & 1.73 & 1.39 & 1.72 \\
\midrule
Mean(Amount Saved) & \$19,336 & \$23,373 & \$19,876 & \$23,648 \\
\bottomrule
\end{tabular}
\begin{tablenotes}
\small
\item Notes: *** p$<$0.01, ** p$<$0.05, * p$<$0.1; Robust standard errors clustered by account and appraiser; Observation counts decline in columns 3 and 4 due to the omission of hearings involving only one ARB member. 
\end{tablenotes}
\end{threeparttable}
\end{table}

\begin{table}[htpb!]
\small
\centering
\caption{Effect of ARB Gender on Home Value Reduction \textit{(DV = Log(AmtSaved+1)} by (Gemini-detected = Regex-based) Appellant Gender}
\label{tab:baseline_gemini_regex_amt}
\begin{threeparttable}
\begin{tabular}{lcccc}
\toprule
Appellant Gender & \textbf{Female} & \textbf{Male} & \textbf{Female} & \textbf{Male} \\
\midrule
  \multicolumn{5}{c}{\textbf{Tax Year, Home Value Bucket \& Appraiser FEs}} \\
\midrule
\rowcolor{Gray}
Female Panel Chair & -0.181** (0.078) & -0.016 (0.052) &  &  \\
\rowcolor{Gray}
Female Panelists (1) &  &  & -0.263** (0.110) & 0.017 (0.085) \\
\rowcolor{Gray}
Female Panelists (2) &  &  & -0.388*** (0.121) & 0.075 (0.085) \\
\rowcolor{Gray}
Female Panelists (3) &  &  & -0.415*** (0.128) & -0.008 (0.093) \\
Online & -0.542*** (0.141) & -0.238** (0.109) & -0.490*** (0.169) & -0.159 (0.118) \\
Market Equity & 1.439*** (0.185) & 1.459*** (0.099) & 1.453*** (0.185) & 1.462*** (0.099) \\
Unequal Appraisal & 0.202** (0.080) & 0.227*** (0.056) & 0.214*** (0.081) & 0.223*** (0.056) \\
Count Prior Protests & -0.566*** (0.051) & -0.531*** (0.033) & -0.591*** (0.055) & -0.557*** (0.034) \\
Saved Last Protest & 0.806*** (0.099) & 0.715*** (0.067) & 0.767*** (0.103) & 0.741*** (0.065) \\
\midrule
Num.Obs. & 17,077 & 45,347 & 15,577 & 42,497 \\
R2 & 0.092 & 0.071 & 0.089 & 0.068 \\
R2 Adj. & 0.073 & 0.063 & 0.068 & 0.059 \\
RMSE & 4.44 & 4.48 & 4.41 & 4.45 \\
\midrule
  \multicolumn{5}{c}{\textbf{Tax Year, Account and Appraiser FEs}} \\
\midrule
\rowcolor{Gray}
Female Panel Chair & -0.387** (0.191) & 0.013 (0.093) &  &  \\
\rowcolor{Gray}
Female Panelists (1) &  &  & -0.332 (0.306) & 0.099 (0.161) \\
\rowcolor{Gray}
Female Panelists (2) &  &  & -0.572* (0.306) & 0.163 (0.153) \\
\rowcolor{Gray}
Female Panelists (3) &  &  & -0.484 (0.354) & 0.137 (0.183) \\
Online & 0.924* (0.475) & 0.209 (0.237) & 1.031* (0.535) & 0.188 (0.273) \\
Market Equity & 0.261 (0.461) & 0.668*** (0.237) & 0.287 (0.476) & 0.698*** (0.239) \\
Unequal Appraisal & 0.191 (0.271) & 0.398*** (0.115) & 0.205 (0.278) & 0.399*** (0.115) \\
Count Prior Protests & -0.407*** (0.153) & -0.190** (0.078) & -0.314* (0.168) & -0.166* (0.085) \\
Saved Last Protest & -2.353*** (0.194) & -2.657*** (0.099) & -2.447*** (0.209) & -2.700*** (0.103) \\
\midrule
Num.Obs. & 17,077 & 45,347 & 15,577 & 42,497 \\
R2 & 0.923 & 0.865 & 0.924 & 0.864 \\
R2 Adj. & 0.230 & 0.262 & 0.226 & 0.252 \\
RMSE & 1.29 & 1.71 & 1.27 & 1.70 \\
\midrule
Mean(Amount Saved) & \$19,788 & \$24,291 & \$20,291 & \$24,540 \\
\bottomrule
\end{tabular}
\begin{tablenotes}
\small
\item Notes: *** p$<$0.01, ** p$<$0.05, * p$<$0.1; Robust standard errors clustered by account and appraiser; Observation counts decline in columns 3 and 4 due to the omission of hearings involving only one ARB member. 
\end{tablenotes}
\end{threeparttable}
\end{table}

Now, perhaps due to social norms, whatever behavior and tones may be exhibited by appellants, they may also be perceived differentially by ARB panelists \citep{cortes2024should}, potentially in ways that also depend on ARB panelists' genders. That is, female ARB panelists may perceive female appellants' behavior differently than male ARB panelists do, conditional on the behavior exhibited. We consider this possibility next, in two ways. First, we attempt to understand whether and to what degree appellants' behavior differentially correlates with hearing outcomes, depending on their gender and that of the ARB panel. Second, we repeat our analysis of the impact of gender (dis)concordance on hearing outcomes, this time conditioning on appellants' behavior within their hearing.

For the first analysis, we conduct four subsample regressions, splitting on combinations of appellant gender and ARB chair gender, to explore correlations; that is, the estimates from these regressions do not necessarily represent causal effects. We focus on the \textit{AnySaved} outcome in these analyses, for the sake of simplicity. We also focus strictly on cases where Gemini's gender predictions align with our regular-expression-based gender predictions to minimize potential measurement error. Thus, we work with approximately three-quarters of the audio-recorded sample (roughly 62,000 records across the four sample splits). Our results are shown in Table~\ref{tab:behavior_impact_by_gender_pair}, where we see a few interesting associations. For example, we see that presenting formal evidence has a significant, positive relationship with home value reductions, and this effect is relatively homogeneous across all gender-pair sub-samples, increasing the probability of a value reduction by approximately 6-7 pp. Further, we observe that 'sounding confident' has a positive impact on hearing outcomes specifically when ARB panels are chaired by men, and that male appellants, in particular, are penalized for disagreeing verbally with the panel. We also observe that `sounding hostile' has a significant negative impact on hearing outcomes, specifically when ARB panels are chaired by women. Importantly, however, we see no effects that would suggest an explanation for why females fare worse (better) in front of female (male) ARB chairs. A particularly salient takeaway from these estimates is that, despite our observation that males are systematically more aggressive in their behavior in front of female ARB chairs (raising their voice, interrupting, disagreeing, sounding hostile), this has very limited material consequences for their hearing outcomes. This suggests that the behavior is tolerated and perhaps even expected. 

\begin{table}[htpb!]
\small
\centering
\caption{Effect of Appellant Behavior by Gender Composition of Appellant and Panel Chair on Home Value Reduction \textit{(DV = AnySaved}}
\label{tab:behavior_impact_by_gender_pair}
\begin{tabular}{lcccc}
\toprule
\textbf{Appellant-Chair Gender} & \textbf{Female-Female} & \textbf{Male-Female} & \textbf{Female-Male} & \textbf{Male-Male} \\
\midrule
App Raises Voice & 0.016 (0.032) & -0.005 (0.017) & 0.021 (0.029) & 0.033* (0.020) \\
App Disagrees & -0.015 (0.014) & -0.029*** (0.009) & -0.023 (0.015) & -0.022** (0.011) \\
App Interrupts & -0.010 (0.013) & 0.012 (0.008) & -0.007 (0.003) & 0.004 (0.009) \\
App Refers to Evidence & 0.069*** (0.011) & 0.067*** (0.007) & 0.073*** (0.012) & 0.070*** (0.008) \\
App Sounds Confident & 0.013 (0.017) & 0.012 (0.012) & 0.040** (0.017) & 0.028** (0.012) \\
App Sounds Frustrated & 0.018 (0.018) & -0.016* (0.010) & 0.012 (0.017) & -0.010 (0.011) \\
App Sounds Defensive & 0.036*** (0.013) & -0.001 (0.009) & 0.007 (0.014) & -0.001 (0.011) \\
%App Sounds Assertive & 0.013 (0.024) & 0.023* (0.013) & -0.001 (0.028) & 0.049*** (0.016) \\
App Sounds Nervous & -0.009 (0.027) & 0.003 (0.025) & 0.025 (0.027) & 0.057* (0.030) \\
App Sounds Angry & 0.041 (0.042) & 0.023 (0.026) & 0.014 (0.039) & -0.024 (0.031) \\
App Sounds Hostile & -0.071* (0.040) & -0.041* (0.021) & -0.022 (0.040) & -0.015 (0.030) \\
App Asks Questions & -0.006 (0.012) & 0.015* (0.008) & 0.003 (0.014) & 0.007 (0.008) \\
App Asks Repeat & -0.009 (0.012) & -0.002 (0.007) & -0.009 (0.013) & -0.016* (0.009) \\
App Asks Speak Louder & -0.005 (0.024) & 0.020 (0.014) & -0.022 (0.025) & -0.010 (0.017) \\
\midrule
Num. Obs. & 9,278 & 24,752 & 7,799 & 20,595 \\
R2 & 0.006 & 0.006 & 0.008 & 0.007 \\
R2 Adj. & 0.005 & 0.005 & 0.006 & 0.006 \\
RMSE & 0.48 & 0.47 & 0.47 & 0.46 \\
\midrule
Mean(Any Saved) & 0.642 & 0.678 & 0.659 & 0.680 \\
\bottomrule
\end{tabular}
\begin{tablenotes}
\small
\item Notes: *** p$<$0.01, ** p$<$0.05, * p$<$0.1; Robust standard errors in parentheses clustered by account and appraiser; Gender compositions correspond to appellant gender (where Gemini coding agreed with regular-expression-based prediction) followed by panel chair gender (e.g., Female-Female indicates both appellant and panel chair are female).
\end{tablenotes}
\end{table}

We now incorporate controls for our Gemini-coded features into our regression specification, focusing on the same outcome measures. As before, we split our sample by appellant gender, incorporating the same set of controls, e.g., measures of whether the hearing is conducted online, the count of prior protest hearings for the property account during our larger sample period, an indicator of whether the most recent hearing, should it exist, resulted in savings, and fixed effects for the home value bucket (or account), the appraiser, and the tax year. Here, we focus on the sample where Gemini's gender predictions agree with the regular-expression-based predictions, the goal being to minimize any role of measurement error. The results of these regressions are reported in Tables~\ref{tab:audio_regs_any} and \ref{tab:audio_regs_amt}. 

\begin{table}[htpb!]
\small
\centering
\caption{Effect of ARB on Home Value Reduction \textit{(DV = AnySaved)} by (Gemini-detected = Regex-based) Appellant Gender}
\label{tab:audio_regs_any}
\begin{threeparttable}
\begin{tabular}{lcccc}
\toprule
\textbf{Appellant Gender} & \textbf{Female} & \textbf{Male} & \textbf{Female} & \textbf{Male} \\
\midrule
\rowcolor{Gray}
Female Panel Chair & -0.015* (0.008) & 0.000 (0.005) & & \\
\rowcolor{Gray}
Female Panelists (1) &  &  & -0.026** (0.012) & 0.002 (0.008) \\
\rowcolor{Gray}
Female Panelists (2) &  &  & -0.036*** (0.013) & 0.010 (0.008) \\
\rowcolor{Gray}
Female Panelists (3) &  &  & -0.039*** (0.013) & 0.003 (0.009) \\
Market Equity & 0.126*** (0.019) & 0.117*** (0.010) & 0.128*** (0.019) & 0.118*** (0.010) \\
Unequal Appraisal & 0.034*** (0.008) & 0.038*** (0.005) & 0.036*** (0.008) & 0.038*** (0.005) \\
Online & -0.046*** (0.015) & -0.008 (0.011) & -0.041** (0.017) & 0.000 (0.012) \\
Count Prior Protests & -0.059*** (0.005) & -0.053*** (0.003) & -0.061*** (0.006) & -0.056*** (0.003) \\
Saved Last Protest & 0.073*** (0.011) & 0.066*** (0.007) & 0.070*** (0.011) & 0.069*** (0.006) \\
\midrule
\multicolumn{5}{l}{\textbf{Gemini-coded Audio Features}} \\
\midrule
App Raises Voice & 0.004 (0.020) & 0.007 (0.013) & 0.013 (0.021) & 0.008 (0.013) \\
App Disagrees & -0.015 (0.010) & -0.023*** (0.007) & -0.013 (0.011) & -0.019*** (0.007) \\
App Interrupts & -0.009 (0.009) & 0.008 (0.006) & -0.008 (0.010) & 0.008 (0.007) \\
App Refers to Evidence & 0.067*** (0.008) & 0.071*** (0.006) & 0.062*** (0.008) & 0.066*** (0.006) \\
App Sounds Confident & 0.027** (0.012) & 0.021** (0.008) & 0.020 (0.013) & 0.016** (0.008) \\
App Sounds Frustrated & 0.022* (0.013) & -0.004 (0.007) & 0.024* (0.013) & -0.001 (0.007) \\
App Sounds Defensive & 0.009 (0.009) & -0.007 (0.007) & 0.005 (0.009) & -0.011 (0.007) \\
%App Sounds Assertive & -0.001 (0.017) & 0.036*** (0.009) & -0.003 (0.018) & 0.036*** (0.010) \\
App Sounds Nervous & 0.006 (0.019) & 0.027 (0.019) & -0.005 (0.019) & 0.028 (0.019) \\
App Sounds Angry & 0.025 (0.029) & -0.002 (0.019) & 0.021 (0.029) & -0.006 (0.019) \\
App Sounds Hostile & -0.034 (0.030) & -0.019 (0.017) & -0.038 (0.030) & -0.014 (0.017) \\
App Asks Questions & -0.004 (0.009) & 0.009 (0.006) & 0.006 (0.009) & 0.012* (0.006) \\
App Asks Repeat & -0.004 (0.009) & -0.005 (0.006) & -0.004 (0.009) & -0.005 (0.006) \\
App Asks Speak Louder & -0.004 (0.016) & 0.011 (0.011) & -0.007 (0.017) & 0.011 (0.011) \\
\midrule
Num. Obs. & 17,077 & 45,347 & 15,577 & 42,497 \\
R2 & 0.085 & 0.064 & 0.080 & 0.059 \\
R2 Adj. & 0.065 & 0.056 & 0.058 & 0.050 \\
RMSE & 0.46 & 0.45 & 0.45 & 0.45 \\
\midrule
Mean(Any Saved) & 0.650 & 0.679 & 0.665 & 0.689 \\
\bottomrule
\end{tabular}
\begin{tablenotes}
\small
\item Notes: *** p$<$0.01, ** p$<$0.05, * p$<$0.1; Robust standard errors in parentheses clustered by account and appraiser; Panelist genders are genders derived from names on hearing records; Appellant genders reflect genders derived from names on hearing records and inferred by Google Gemini on hearing audio; Observation counts decline in columns 3 and 4 due to the omission of hearings involving only one ARB panelist; Estimations incorporate home value bucket FEs.
\end{tablenotes}
\end{threeparttable}
\end{table}

\begin{table}[htpb!]

\small
\centering
\caption{Effect of ARB on Home Value Reduction \textit{(DV = Log(AmtSaved+1)} by (Gemini-detected = Regex-based) Appellant Gender}
\label{tab:audio_regs_amt}
\begin{threeparttable}
\begin{tabular}{lcccc}
\toprule
\textbf{Appellant Gender} & \textbf{Female} & \textbf{Male} & \textbf{Female} & \textbf{Male} \\
\midrule
\rowcolor{Gray}
Female Panel Chair & -0.169** (0.076) & -0.018 (0.051) &  &  \\
\rowcolor{Gray}
Female Panelists (1) &  &  & -0.250** (0.110) & 0.011 (0.083) \\
\rowcolor{Gray}
Female Panelists (2) &  &  & -0.368*** (0.120) & 0.075 (0.084) \\
\rowcolor{Gray}
Female Panelists (3) &  &  & -0.388*** (0.127) & -0.006 (0.092) \\
Market Equity & 1.426*** (0.182) & 1.476*** (0.101) & 1.448*** (0.182) & 1.481*** (0.101) \\
Unequal Appraisal & 0.194** (0.080) & 0.220*** (0.056) & 0.213*** (0.080) & 0.218*** (0.056) \\
Online & -0.443*** (0.140) & -0.199* (0.108) & -0.396** (0.166) & -0.132 (0.117) \\
Count Prior Protests & -0.587*** (0.052) & -0.543*** (0.033) & -0.609*** (0.056) & -0.567*** (0.034) \\
Saved Last Protest & 0.752*** (0.100) & 0.694*** (0.067) & 0.711*** (0.103) & 0.716*** (0.065) \\
\midrule
\multicolumn{5}{l}{\textbf{Gemini-coded Audio Features}} \\
\midrule
App Raises Voice & 0.062 (0.196) & 0.071 (0.129) & 0.149 (0.203) & 0.070 (0.129) \\
App Disagrees & -0.164* (0.097) & -0.276*** (0.067) & -0.141 (0.105) & -0.245*** (0.074) \\
App Interrupts & -0.131 (0.093) & 0.054 (0.062) & -0.124 (0.095) & 0.048 (0.064) \\
App Refers to Evidence & 0.856*** (0.074) & 0.880*** (0.054) & 0.814*** (0.078) & 0.836*** (0.053) \\
App Sounds Confident & 0.255** (0.118) & 0.247** (0.076) & 0.184 (0.127) & 0.199** (0.077) \\
App Sounds Frustrated & 0.332*** (0.119) & 0.074 (0.070) & 0.349*** (0.123) & 0.105 (0.072) \\
App Sounds Defensive & 0.109 (0.089) & -0.060 (0.066) & 0.073 (0.091) & -0.103 (0.067) \\
%App Sounds Assertive & 0.019 (0.167) & 0.368*** (0.091) & 0.011 (0.175) & 0.369*** (0.097) \\
App Sounds Nervous & 0.144 (0.191) & 0.254 (0.183) & 0.011 (0.190) & 0.252 (0.183) \\
App Sounds Angry & 0.215 (0.286) & 0.069 (0.190) & 0.195 (0.286) & 0.052 (0.193) \\
App Sounds Hostile & -0.344 (0.297) & -0.277 (0.171) & -0.387 (0.292) & -0.247 (0.173) \\
App Asks Questions & -0.011 (0.086) & 0.169*** (0.062) & 0.077 (0.092) & 0.202*** (0.063) \\
App Asks Repeat & 0.007 (0.087) & -0.020 (0.055) & 0.006 (0.092) & -0.004 (0.056) \\
App Asks Speak Louder & -0.068 (0.160) & 0.080 (0.102) & -0.102 (0.168) & 0.066 (0.107) \\
\midrule
Num. Obs. & 17,077 & 45,347 & 15,577 & 42,497 \\
R2 & 0.102 & 0.080 & 0.097 & 0.076 \\
R2 Adj. & 0.082 & 0.072 & 0.076 & 0.067 \\
RMSE & 4.42 & 4.45 & 4.39 & 4.43 \\
\midrule
Mean(Amount Saved) & \$19,788 & \$24,291 & \$20,291 & \$24,540 \\
\bottomrule
\end{tabular}
\begin{tablenotes}
\small
\item Notes: *** p$<$0.01, ** p$<$0.05, * p$<$0.1; Robust standard errors in parentheses clustered by account and appraiser; Panelist genders are genders derived from names on hearing records; Appellant genders reflect genders derived from names on hearing records and inferred by Google Gemini on hearing audio; Observation counts decline in columns 3 and 4 due to the omission of hearings involving only one ARB panelist; Estimations incorporate home value bucket FEs.
\end{tablenotes}
\end{threeparttable}
\end{table}

In all-female homeowner models, we again observe a negative and significant gender concordance effect. The point estimates we observe are roughly equivalent to those observed in our baseline models, reported in Tables~\ref{tab:baseline_any} and \ref{tab:baseline_amt}, which suggests that appellant behaviors and tones do not mediate the gender-concordance effect.

Finally, we repeat the estimation accounting for high-dimensional embedded representations of the broader, unstructured content of hearing audio recordings, rather than conditioning upon specific Gemini-coded features. We achieve this employing a double machine learning framework \citep{chernozhukov2018double}, an approach that has been taken in other recent literature to account for unstructured data in a relatively data-driven, non-parametric manner \citep{manzoor2024influence}. The double machine learning framework is useful here because it simultaneously enables us to accommodate many covariates, i.e., our embedding dimensions, and their likely non-linear interactive relationships with appellant gender, panelist gender, and hearing outcomes. We modeled the relationships between our nuisance parameters (structured covariates and text-embedding dimensions), the outcome, and the ARB panelist gender via a pair of random forest models. The text-embedding for each hearing audio recording was obtained by first passing the recording through Open AI's Whisper Medium speech-to-text model. Because Whisper model inference is relatively slow (compared to inference with Gemini Flash, at least), we conduct this exercise focusing strictly on hearing audio records from 2022. The textual transcription for each hearing recording was then passed through Google's `text-embedding-004' model, yielding a 768-dimensional vector representation of the spoken content.\footnote{Stack Overflow Blog | ``An Intuitive Introduction to Text Embeddings'': \href{https://stackoverflow.blog/2023/11/09/an-intuitive-introduction-to-text-embeddings/}{https://stackoverflow.blog/2023/11/09/an-intuitive-introduction-to-text-embeddings/}} Finally, each embedding dimension is treated as a distinct covariate in our estimation, which we implement employing the \textit{DoubleML} package in \textit{R}, specifically its implementation of a partial linear specification. We focus on the linear probability model that considers the \textit{Any Saved} outcome. The results of these estimations are presented in Table~\ref{tab:double_ml_any}, where we again observe consistent results.

\begin{table}[ht]
\small
\centering
\caption{Double Machine Learning Estimation of the Effect of ARB Chair Gender on Any Home Value Reduction, by Appellant Gender, Conditional on Hearing Transcript Embedding}
\begin{threeparttable}
\begin{tabular}{lcc}
\hline
Appellant Gender & \textbf{Female} & \textbf{Male} \\ 
\hline
Female Panel Chair & -0.027* (0.016) & -0.002 (0.012) \\
\hline
Num.Obs. & 3,731 & 7,266 \\
\midrule
Mean(Any Saved) & 0.510 & 0.518 \\
\bottomrule
\end{tabular}
\label{tab:double_ml_any}
\begin{tablenotes}
\small
\item Notes: *** p$<$0.01, ** p$<$0.05, * p$<$0.1; Appellant and panelist genders are derived from names on hearing records; Estimations account for home value bucket FEs, the same set of controls and hearing transcript embeddings; Estimation performed on hearings conducted for tax year 2022.
\end{tablenotes}
\end{threeparttable}
\end{table}

That the gender concordance effects persist for female homeowners across both approaches, conditioning on hearing activity, suggests that, although appellants' (or panelists') behavior may contribute to differential hearing outcomes, other unmeasured (unvoiced) factors continue to play a significant role. 

\section{Discussion}
\label{sec:discussion}

Our study provides new insights into the role of gender concordance in property tax protest hearings, revealing significant biases that disproportionately disadvantage female appellants. We find that female homeowners are systematically less likely to secure reductions in their assessed property values, particularly when they appear before female-dominated appraisal review board (ARB) panels. These gender concordance effects are substantial, with female appellants facing a notably lower probability of success when confronted with female panelists than male panelists.

We further explore the behavioral dynamics underlying these disparities by leveraging unstructured audio data processed through a multi-modal large language model (Gemini 1.5 Flash). Although we find that female homeowners exhibit differences in behavior and tone depending on the gender composition of the ARB panel, these differences are generally modest. Further, controlling for appellant behavior or hearing activity more generally, we continue to find that female appellants are systematically less likely to achieve favorable outcomes when faced by female (versus male) panelists, suggesting that the biases we document are likely rooted in unobserved or implicit factors influencing ARB decision-making.  

Our findings contribute to the growing literature on gender biases in evaluation processes by highlighting a previously undocumented source of bias in the context of property taxation. Our results imply that gender biases in such fundamental administrative processes can exacerbate existing inequalities, with tangible financial consequences for affected individuals. This is particularly concerning given the role of property taxes in funding essential public services, meaning that gender-based disparities in tax burdens can have broad societal implications.

Moreover, our study underscores the importance of considering gender dynamics in seemingly neutral administrative settings and calls for policymakers to reassess the structure and procedures of ARB hearings. Structural adjustments, such as diversifying panel compositions or introducing bias-mitigation training, may help reduce the influence of implicit biases.

Finally, our use of large-scale administrative data combined with unstructured audio analysis demonstrates the potential for advanced AI tools to enhance our understanding of complex social dynamics. The ability to systematically analyze behavioral and tonal cues at scale opens new avenues for research into decision-making processes, especially in contexts where bias may be subtle or difficult to detect using traditional methods. 

Our work is, of course, subject to several limitations. First, we focus on a single geography, Harris, County, Texas. As such, generalizability is a question. Houston may have a unique sociocultural climate, and the local market may be subject to idiosyncrasies in terms of norms and laws, as well as its property appraisal appeals process. That said we have observed broad evidence of gender-concordant bias across the entire pool of ARB panelists (see Figure~\ref{fig:panelist_FEs}), suggesting this phenomenon may be more general, cutting across political and social groups. We can evaluate the generalizability of our results using the SANS framework (selection, attrition, naturalness, and scalability) developed in \citep{List2020}. Regarding selection, our sample is representative of hearings conducted without the involvement of professional agents in Harris County, the target population. However, we note potential selection bias, as females may be less likely to participate in hearings due to anticipated biases against them. Notably, in our setting, participants do not know the gender of the panel members before the hearing. Given the similarity of hearings across Texas's 241 counties, our sample may also be broadly representative of hearings conducted without professional agents across the state. Considering attrition, experiences, and outcomes in prior hearings could influence individuals' willingness to participate in future hearings, potentially affecting the composition of future samples. Regarding naturalness, both homeowners and panelists operate within a natural margin, engaging in real decisions rather than an artificial task designed for research purposes. Regarding scalability, there are no clear reasons to anticipate a voltage drop if the analysis is scaled to other counties in Texas. However, once our findings on biases in this context become public, behavior may change.

In sum, we view our findings as a WAVE1 insight in the nomenclature of \citep{List2020}. Replications are necessary to determine whether our results generalize to other administrative processes, and future work can and should explore these dynamics by leveraging administrative data from other areas of the United States. 

Second, our analysis relies on the accuracy of gender imputations based on first names, which, although generally reliable, may introduce some degree of misclassification. This potential misclassification could slightly influence our estimates of gender effects, though we expect such biases to be minimal given the robustness of our results. Notably, employing gender codings from Gemini's audio annotations rather than appellant names yields consistent results, suggesting that any measurement error here does not explain our findings. Nonetheless, future research could improve upon this by using direct measures of gender where available.

Third, while we leverage a state-of-the-art multi-modal large language model (Gemini 1.5 Flash) to analyze audio recordings, the model’s inferences about tone and behavior are not without error. Although we validate some of these inferences against known ground truths from administrative data, and we also compared these inferences with annotations from human coders for a random sample of recordings (see Appendix B), there remains the possibility that certain nuances of human interaction—especially those influenced by socio-cultural context—are missed or misinterpreted by the model. This limitation suggests that human coding, or a hybrid approach combining machine learning with human validation may be a valuable avenue for future research.

Fourth, while we control for various structured and unstructured factors, there may still be unobserved variables influencing both the behavior of appellants and the decisions of ARB panelists. Our analysis of behavior and tone during hearings focuses primarily on what can be quantified, which may overlook subtler aspects of interpersonal dynamics that could influence outcomes.

Overall, these limitations highlight the need for further research. Despite these limitations, our findings provide compelling evidence of gender (dis)concordance effects in property tax protest procedures, with significant implications for policy and practice. Further, our work highlights the significant potential of multimodal LLMs for processing information from large-scale unstructured administrative data, to better understand procedural processes and outcomes.

\section{Conclusion}
\label{sec:conclusion}

Our study sheds light on a previously under-explored area of gender bias within the context of property tax protest hearings. We document significant gender-concordance effects, where female homeowners face systematic disadvantages in achieving favorable outcomes, particularly when their cases are adjudicated by female ARB panelists. These findings suggest that implicit biases, rather than observable behaviors, play a crucial role in driving these disparities. The implications of our study extend beyond the specific context of property taxation, highlighting the broader risks of gender bias in administrative processes that involve subjective evaluations.

Our results underscore the importance of considering gender dynamics in the design and oversight of evaluative processes, particularly those that impact individuals’ financial well-being. As policymakers and practitioners work to enhance fairness and equity in tax systems and other administrative domains, our findings point to the need for interventions that can mitigate gender-based biases. Future research should continue to investigate these dynamics across different regions and contexts, leveraging diverse methodologies to better understand and address the underlying causes of such biases. By doing so, we can move closer to ensuring that all individuals, regardless of gender, are treated equitably in critical administrative proceedings.

Beyond these contributions, our results could also affect future behavior by appellants and panelists, reducing biases, in which case they would represent a public service. Indeed, when individuals know they are being recorded, their behavior often changes due to the observer effect, where heightened scrutiny influences actions. Teachers, for instance, tend to modify their teaching practices during evaluations, focusing more on structured lessons and student engagement to meet professional standards \citep{Kane2013}. Similarly, judges may exhibit greater formality in court proceedings to align with public and appellate expectations, especially when media coverage is present \citep{Lim2015}. Legislators similarly adjust their rhetoric during televised sessions, emphasizing performative or emotional appeals to resonate with constituents \citep{OSNABRÜGGE_HOBOLT_RODON_2021}. This shift toward transparency and visibility creates both opportunities for accountability and challenges in maintaining authentic behavior. Research consistently shows that such surveillance amplifies the emphasis on appearance and messaging, often at the expense of substance \citep{Ash2017, BOUSSALIS_COAN_HOLMAN_MÜLLER_2021}. Furthermore, our paper illustrates how advancements in AI can enable the analysis of recordings and data previously deemed unsuitable for examination, uncovering insights akin to how DNA revolutionized cold case investigations and helped to solve decades-old mysteries. 

Finally, the dissemination of our results could have both immediate effects on property tax protest hearings in Texas and inspire a redesign of the administrative process in the future. Regarding immediate effects, the county appraisal districts in Texas (recall there are 241 counties), the Texas comptroller overseeing these county appraisal districts, or the panelists themselves could adjust their procedures and behavior to mitigate the biases we document. As for redesigning the administrative process, hearings could incorporate AI consulting bots to prevent biases and assist in adjudication decisions. Indeed, in follow-up work, we are exploring how AI can enhance the adjudication process.

\clearpage
\bibliographystyle{aej}
\bibliography{gender_bias_property_tax}

@article{avenancio2022assessment,
  title={The assessment gap: Racial inequalities in property taxation},
  author={Avenancio-Le{\'o}n, Carlos F and Howard, Troup},
  journal={The Quarterly Journal of Economics},
  volume={137},
  number={3},
  pages={1383--1434},
  year={2022},
  publisher={Oxford University Press}
}

@article{gemini2024,
      title={Gemini 1.5: Unlocking multimodal understanding across millions of tokens of context}, 
      author={Gemini Team},
      year={2024},
      journal={arXiv},
      url={https://arxiv.org/abs/2403.05530} 
}

@inproceedings{lin2024listen,
  title={Listen and Speak Fairly: a Study on Semantic Gender Bias in Speech Integrated Large Language Models},
  author={Lin, Yi-Cheng and Lin, Tzu-Quan and Yang, Chih-Kai and Lu, Ke-Han and Chen, Wei-Chih and Kuan, Chun-Yi and Lee, Hung-yi},
  booktitle={2024 IEEE Spoken Language Technology Workshop (SLT)},
  pages={439--446},
  year={2024},
  organization={IEEE}
}

@article{cohen1960coefficient,
  title={A coefficient of agreement for nominal scales},
  author={Cohen, Jacob},
  journal={Educational and psychological measurement},
  volume={20},
  number={1},
  pages={37--46},
  year={1960},
  publisher={Sage Publications Sage CA: Thousand Oaks, CA}
}

@misc{krippendorff2011computing,
  title={Computing Krippendorff’s alpha-reliability},
  author={Krippendorff, Klaus},
  year={2011},
  publisher={Citeseer}
}

@article{schanke2024digital,
  title={Digital Lyrebirds: Experimental Evidence That Voice-Based Deep Fakes Influence Trust},
  author={Schanke, Scott and Burtch, Gordon and Ray, Gautam},
  journal={Management Science},
  year={2024},
  publisher={INFORMS}
}

@article{Stantcheva2023,
   author = "Stantcheva, Stefanie",
   title = "How to Run Surveys: A Guide to Creating Your Own Identifying Variation and Revealing the Invisible", 
   journal= "Annual Review of Economics",
   year = "2023",
   volume = "15",
   number = "Volume 15, 2023",
   pages = "205-234",
}

@techreport{cortes2024gender,
  title={Gender Differences in Negotiations and Labor Market Outcomes: Evidence from an Information Intervention with College Students},
  author={Cort{\'e}s, Patricia and French, Jacob and Pan, Jessica and Zafar, Basit},
  year={2024},
  institution={National Bureau of Economic Research}
}

@article{cortes2024should,
  title={Should mothers work? how perceptions of the social norm affect individual attitudes toward work in the us},
  author={Cort{\'e}s, Patricia and Ko{\c{s}}ar, Gizem and Pan, Jessica and Zafar, Basit},
  journal={Review of Economics and Statistics},
  pages={1--28},
  year={2024},
  publisher={MIT Press 255 Main Street, 9th Floor, Cambridge, Massachusetts 02142, USA~…}
}

@book{good2013permutation,
  title={Permutation tests: a practical guide to resampling methods for testing hypotheses},
  author={Good, Phillip},
  year={2013},
  publisher={Springer Science \& Business Media}
}

@article{Stantcheva2020,
   author = "Stantcheva, Stefanie",
   title = "Understanding Economic Policies: What do People Know and How Can they Learn?", 
   journal= "Harvard University Working Paper",
   year = "2020",
}

@article{Gentzkow2019,
Author = {Gentzkow, Matthew and Kelly, Bryan and Taddy, Matt},
Title = {Text as Data},
Journal = {Journal of Economic Literature},
Volume = {57},
Number = {3},
Year = {2019},
Month = {September},
Pages = {535‚Äì74},
}

@article{Gentzkow2010,
author = {Gentzkow, Matthew and Shapiro, Jesse M.},
title = {What Drives Media Slant? Evidence From U.S. Daily Newspapers},
journal = {Econometrica},
volume = {78},
number = {1},
pages = {35-71},
year = {2010}
}

@misc{chernozhukov2018double,
  title={Double/debiased machine learning for treatment and structural parameters},
  author={Chernozhukov, Victor and Chetverikov, Denis and Demirer, Mert and Duflo, Esther and Hansen, Christian and Newey, Whitney and Robins, James},
  year={2018},
  publisher={Oxford University Press Oxford, UK}
}

@article{manzoor2024influence,
  title={Influence via ethos: On the persuasive power of reputation in deliberation online},
  author={Manzoor, Emaad and Chen, George H and Lee, Dokyun and Smith, Michael D},
  journal={Management Science},
  volume={70},
  number={3},
  pages={1613--1634},
  year={2024},
  publisher={INFORMS}
}

@article{young2019channeling,
  title={Channeling fisher: Randomization tests and the statistical insignificance of seemingly significant experimental results},
  author={Young, Alwyn},
  journal={The Quarterly Journal of Economics},
  volume={134},
  number={2},
  pages={557--598},
  year={2019},
  publisher={Oxford University Press}
}

@article{Nathan2024,
    title = {{Paying Your Fair Share: Perceived Fairness and Tax Compliance}},
    year = {2024},
    journal = {NBER. Working Paper 32588},
    author = {Nathan, Brad and Perez-Truglia, Ricardo and Zentner, Alejandro}
}

@ARTICLE{Nathan2022,
title = {Is the Partisan Divide Real? Polarization in Preferences for Redistribution},
author = {Nathan, Brad and Perez-Truglia, Ricardo and Zentner, Alejandro},
year = {2022},
journal = {AEA Papers and Proceedings},
volume = {112},
pages = {156-62},
}

@article{AndrewJustin2023,
    title = {Racial Inequality in Property Tax Appeals: Evidence from Field Experiments with Homeowners and Assessors},
    year = {2023},
    journal = {Working Paper},
    author = {Holz, Justin and Novgorodsky, David and Simon, Andrew}
}

@article{bordalo2016stereotypes,
  title={Stereotypes},
  author={Bordalo, Pedro and Coffman, Katherine and Gennaioli, Nicola and Shleifer, Andrei},
  journal={The Quarterly Journal of Economics},
  volume={131},
  number={4},
  pages={1753--1794},
  year={2016},
  publisher={MIT Press}
}

@inbook{tajfel1979integrative,
  author={Tajfel, H. and Turner, J. C.},
  title={An integrative theory of intergroup conflict \textit{In:} `The Social Psychology of Intergroup Relations'},
  pages = {56--65},
  year = 1979,
  editor = {Hatch, Mary Jo and Schultz, Majken},
  publisher={Monterey, CA: Brooks/Cole.},
}

@article{cullen2023old,
  title={The old boys’ club: Schmoozing and the gender gap},
  author={Cullen, Zo{\"e} and Perez-Truglia, Ricardo},
  journal={American Economic Review},
  volume={113},
  number={7},
  pages={1703--1740},
  year={2023},
  publisher={American Economic Association 2014 Broadway, Suite 305, Nashville, TN 37203}
}

@article{recalde2023gender,
  title={Gender differences in negotiation: can interventions reduce the gap?},
  author={Recalde, Maria P and Vesterlund, Lise},
  journal={Annual Review of Economics},
  volume={15},
  number={1},
  pages={633--657},
  year={2023},
  publisher={Annual Reviews}
}

@article{ely1994effects,
  title={The effects of organizational demographics and social identity on relationships among professional women},
  author={Ely, Robin J},
  journal={Administrative science quarterly},
  pages={203--238},
  year={1994},
  publisher={JSTOR}
}

@article{bowles2007social,
  title={Social incentives for gender differences in the propensity to initiate negotiations: Sometimes it does hurt to ask},
  author={Bowles, Hannah Riley and Babcock, Linda and Lai, Lei},
  journal={Organizational Behavior and human decision Processes},
  volume={103},
  number={1},
  pages={84--103},
  year={2007},
  publisher={Elsevier}
}

@article{coffman2014evidence,
  title={Evidence on self-stereotyping and the contribution of ideas},
  author={Coffman, Katherine Baldiga},
  journal={The Quarterly Journal of Economics},
  volume={129},
  number={4},
  pages={1625--1660},
  year={2014},
  publisher={MIT Press}
}

@article{greenberg2017activist,
  title={Activist choice homophily and the crowdfunding of female founders},
  author={Greenberg, Jason and Mollick, Ethan},
  journal={Administrative Science Quarterly},
  volume={62},
  number={2},
  pages={341--374},
  year={2017},
  publisher={Sage Publications Sage CA: Los Angeles, CA}
}

@article{knepper2018shadow,
  title={When the shadow is the substance: Judge gender and the outcomes of workplace sex discrimination cases},
  author={Knepper, Matthew},
  journal={Journal of Labor Economics},
  volume={36},
  number={3},
  pages={623--664},
  year={2018},
  publisher={University of Chicago Press Chicago, IL}
}

@article{coffman2021role,
  title={The role of beliefs in driving gender discrimination},
  author={Coffman, Katherine and Exley, Christine and Niederle, Muriel},
  journal={Management Science},
  volume={67},
  number={6},
  pages={3551--3569},
  year={2021},
  publisher={INFORMS}
}

@article{rosenbaum1985constructing,
  title={Constructing a control group using multivariate matched sampling methods that incorporate the propensity score},
  author={Rosenbaum, Paul R and Rubin, Donald B},
  journal={The American Statistician},
  volume={39},
  number={1},
  pages={33--38},
  year={1985},
  publisher={Taylor \& Francis}
}

@article{austin2009balance,
  title={Balance diagnostics for comparing the distribution of baseline covariates between treatment groups in propensity-score matched samples},
  author={Austin, Peter C},
  journal={Statistics in medicine},
  volume={28},
  number={25},
  pages={3083--3107},
  year={2009},
  publisher={Wiley Online Library}
}

@techreport{cabral2012hated,
  title={The hated property tax: salience, tax rates, and tax revolts},
  author={Cabral, Marika and Hoxby, Caroline},
  year={2012},
  institution={National Bureau of Economic Research}
}

@article{KRESCH2023,
title = {Sanitation and property tax compliance: Analyzing the social contract in Brazil},
journal = {Journal of Development Economics},
volume = {160},
year = {2023},
author = {Evan Plous Kresch and Mark Walker and Michael Carlos Best and Francois Gerard and Joana Naritomi},
}

@article{brock2023discriminatory,
  title={Discriminatory lending: Evidence from bankers in the lab},
  author={Brock, J Michelle and De Haas, Ralph},
  journal={American Economic Journal: Applied Economics},
  volume={15},
  number={2},
  pages={31--68},
  year={2023},
  publisher={American Economic Association 2014 Broadway, Suite 305, Nashville, TN 37203-2425}
}

@article{Kane2013,
  author    = {Thomas J. Kane and Douglas O. Staiger and Daniel F. McCaffrey and Thomas Miller},
  title     = {Have we identified effective teachers? Validating measures of effective teaching using random assignment},
  year      = {2013},
  journal   = {MET Project Research Paper},
  publisher = {Bill & Melinda Gates Foundation},
}

@article{Lim2015,
Author = {Lim, Claire S. H. and Snyder, James M., Jr. and Strömberg, David},
Title = {The Judge, the Politician, and the Press: Newspaper Coverage and Criminal Sentencing across Electoral Systems},
Journal = {American Economic Journal: Applied Economics},
Volume = {7},
Number = {4},
Year = {2015},
Month = {October},
Pages = {103–35},
}

@article{Ash2017,
  author    = {Elliot Ash and W. Bentley MacLeod},
  title     = {Transparency and Emotionality in Political Speech},
  year      = {2017},
  journal   = {American Economic Journal: Microeconomics},
  volume    = {9},
  number    = {1},
  pages     = {1-30},
}

@article{OSNABRÜGGE_HOBOLT_RODON_2021, 
title={Playing to the Gallery: Emotive Rhetoric in Parliaments}, 
volume={115},
number={3}, 
journal={American Political Science Review}, author={Osnabrugge, Moritz and Hobolt, Sara B. and Rodon, Toni}, 
year={2021}, 
pages={885–899},
}

@article{BOUSSALIS_COAN_HOLMAN_MÜLLER_2021, title={Gender, Candidate Emotional Expression, and Voter Reactions During Televised Debates}, volume={115}, 
number={4},
journal={American Political Science Review},
author={Boussalis, Constantine and Coan, Travis G. and Holman, Mirya R. and MÜller, Stefan}, year={2021},
pages={1242–1257}
}

@article{boyd2017effects,
  title={The effects of trial judge gender and public opinion on criminal sentencing decisions},
  author={Boyd, Christina L and Nelson, Michael J},
  journal={Vand. L. Rev.},
  volume={70},
  pages={1819},
  year={2017},
  publisher={HeinOnline}
}

@article{lau2021does,
  title={Does patient-physician gender concordance influence patient perceptions or outcomes?},
  author={Lau, Emily S and Hayes, Sharonne N and Volgman, Annabelle Santos and Lindley, Kathryn and Pepine, Carl J and Wood, Malissa J and American College of Cardiology Cardiovascular Disease in Women Section},
  journal={Journal of the American College of Cardiology},
  volume={77},
  number={8},
  pages={1135--1138},
  year={2021},
  publisher={American College of Cardiology Foundation Washington DC}
}

@article{kaplan2021predictrace,
  title={Predictrace: Predict the race and gender of a given name using census and social security administration data},
  author={Kaplan, J},
  journal={R package version},
  volume={2},
  number={0},
  year={2021}
}

@article{cabral2024gender,
  title={Gender Differences in Medical Evaluations: Evidence from Randomly Assigned Doctors},
  author={Cabral, Marika and Dillender, Marcus},
  journal={American Economic Review},
  volume={114},
  number={2},
  pages={462--499},
  year={2024},
  publisher={American Economic Association 2014 Broadway, Suite 305, Nashville, TN 37203}
}

@techreport{giaccobasso2022my,
  title={Where do my tax dollars go? Tax morale effects of perceived government spending},
  author={Giaccobasso, Matias and Nathan, Brad C and Perez-Truglia, Ricardo and Zentner, Alejandro},
  year={2025},
  institution={American Economic Journal: Applied Economics. Forthcoming}
}

@article{takeshita2020association,
  title={Association of racial/ethnic and gender concordance between patients and physicians with patient experience ratings},
  author={Takeshita, Junko and Wang, Shiyu and Loren, Alison W and Mitra, Nandita and Shults, Justine and Shin, Daniel B and Sawinski, Deirdre L},
  journal={JAMA network open},
  volume={3},
  number={11},
  pages={e2024583--e2024583},
  year={2020},
  publisher={American Medical Association}
}

@article{List2020,
    title = {{Non est Disputandum de Generalizability? A Glimpse into The External Validity Trial}},
    year = {2020},
    journal = {NBER Working Paper No. 27535},
    author = {List, John A.}
}

@article{nathan2020my,
Author = {Nathan, Brad and Perez-Truglia, Ricardo and Zentner, Alejandro},
Title = {My Taxes Are Too Darn High: Why Do Households Protest Their Taxes?},
Journal = {American Economic Journal: Economic Policy},
Volume = {17},
Number = {1},
Year = {2025},
Month = {February},
Pages = {273–310},
}

\clearpage

\section*{APPENDIX A: Audio Annotation with Google Gemini 1.5 Flash}

For each audio recording, we obtain annotations using Google's `genai' API (Application Programming Interface), using Python. All API requests employed `gemini-1.5-flash-001' and requested a JSON formatted response. Audio annotations for hearings from 2022 were obtained in late August of 2024. Additional hearing records were obtained for 2013 through 2021 in early October of 2024 and were coded in the latter half of that month. For every recording, a zero-shot prompt was employed, using the default temperature setting (a value of 1.0). The exact prompt for the coding exercise was as follows.

\begin{verbatim}

The following audio file is a recording of an appraisal review board hearing 
in Houston. Listen carefully and answer the following questions. Be sure to 
use exactly one of the provided response options for each question, when 
indicated, and never respond with 'null'. Please analyze the following audio 
file and respond with the answers to each question in the specified 
JSON format. Ensure that the JSON keys correspond exactly to the questions 
asked. The JSON structure should look like this:

{
  "appellant_who_speaks_most_gender": "ANSWER WITH MALE or FEMALE",
  "appellant_property_owner_or_hired_agent": "ANSWER WITH OWNER or AGENT",
  "appellant_attending_online": "ANSWER YES or NO",
  "appellant_asks_board_member_or_appraiser_to_repeat": "ANSWER YES or NO",
  "appellant_asks_board_member_or_appraiser_to_speak_louder": 
  "ANSWER YES or NO",
  "appellant_ever_sounds_accusatory_or_hostile": ANSWER YES or NO",
  "appellant_sounds_defensive": "ANSWER YES or NO", 
  "appellant_asks_board_member_or_appraiser_any_questions": 
  "ANSWER YES or NO",
  "appellant_presents_formal_evidence": "ANSWER YES or NO",
  "appellant_ever_raises_voice": "ANSWER WITH YES or NO",
  "appellant_sounds_frustrated": "ANSWER WITH YES or NO",
  "appellant_sounds_annoyed": "ANSWER WITH YES or NO",
  "appellant_sounds_nervous": "ANSWER WITH YES or NO",
  "appellant_sounds_confident": "ANSWER WITH YES or NO",
  "appellant_sounds_angry": "ANSWER WITH YES or NO",
  "appellant_ever_disagrees_with_board_member_or_appraiser": 
  "ANSWER WITH YES or NO",
  "appellant_ever_interrupts_board_member_or_appraiser": 
  "ANSWER WITH YES or NO",
  "board_member_or_appraiser_ask_appellant_any_questions": 
  "ANSWER WITH YES or NO",
  "board_member_or_appraiser_ever_asks_appellant_to_repeat": 
  "ANSWER WITH YES or NO",
  "board_member_or_appraiser_ever_ask_appellant_to_speak_louder": 
  "ANSWER WITH YES or NO",
  "board_member_or_appraiser_sound_accusatory_or_hostile": 
  "ANSWER WITH YES or NO",
  "board_member_or_appraiser_sound_defensive": "ANSWER WITH YES or NO",
  "board_member_or_appraiser_sound_confident": "ANSWER WITH YES or NO",
  "board_member_or_appraiser_sound_nervous": "ANSWER WITH YES or NO",
  "board_member_or_appraiser_ever_raise_voice": "ANSWER WITH YES or NO",
  "board_member_or_appraiser_sound_frustrated": "ANSWER WITH YES or NO",
  "board_member_or_appraiser_sound_annoyed": "ANSWER WITH YES or NO",
  "board_member_or_appraiser_sound_angry": "ANSWER WITH YES or NO",
  "board_member_or_appraiser_ever_disagree_with_appellant": 
  "ANSWER WITH YES or NO",
  "board_member_or_appraiser_ever_interrupt_appellant": 
  "ANSWER WITH YES or NO"
}
\end{verbatim}

The resulting JSON responses were then parsed into numeric measures. Although the Gemini API allows users to configure requests to produce JSON formatted responses, the formatting of responses is not guaranteed with Flash (it \textit{is} guaranteed with Gemini 1.5 Pro, but inference with Gemini Pro is much slower, and an equivalent length request/response costs approximately 20x as much to obtain as from Gemini Flash). Some audio recordings, when annotated, therefore, resulted in malformed responses and were excluded from our sample. 

\clearpage
\section*{APPENDIX B: Comparing M-LLM and Human Audio Annotations}

A natural question is whether the annotations provided by Gemini 1.5 Flash are trustworthy and reflect annotations that one might obtain from a human annotator. To address this question, we undertook two analyses, comparing annotations we obtained from Gemini to those we obtained from human coders. Before considering those results, however, it is important to consider several points. 

First, every item we have annotated is characterized by at least some degree of subjectivity. This is obviously true of tonal questions, but it also applies to items related to appellant or panelist behaviors during hearings, which may seem highly objective at first blush. Consider, for example, the annotation for whether an appellant has asked questions of the panel. A question arises as to whether we should include rhetorical questions or not. Similarly, whether an appellant or panelist has yelled (i.e., `raised their voice') can be a highly subjective determination. For this reason, it is important to recognize not only that we lack ground truth labels for these annotations, but that, to some extent, ground truth might be viewed as altogether undefined. Accordingly, the analyses we conduct here are perhaps best viewed not as a validation exercise but as much as an assessment of whether there is a `meaningful signal' in Gemini's annotation output. Put another way: it is unclear what threshold or standard one might wish to apply here, to judge whether Gemini's annotations are trustworthy.

Second, because we lack ground truth and because there is some degree of arbitrariness in the features we elect to annotate, the analysis we report in Table~\ref{tab:double_ml_any}, wherein we account for high-dimensional embeddings of all hearing content, is quite important to bear in mind. That analysis effectively conditions upon all speech recorded during a hearing from an appellant, the appraiser, and all panelists in a data-driven manner. That we obtain consistent results in that analysis is reassuring, and suggests our findings are not a function of the items we chose to annotate, nor any sources of bias in the annotations themselves. 

Third, and last, on the subject of bias, a great deal of work has documented evidence of gender bias in M-LLMs processing and annotation or response regarding audio input \citep{lin2024listen}. Accordingly, some of our annotations may reflect bias in model training. For example, that males are coded as systematically more likely to raise their voices in front of female ARB chairs may reflect true behavior, or it may reflect a prejudicial evaluation on the part of the M-LLM that derives from biases present in training data, or in the labels generated during model calibration (e.g., based on Reinforcement Learning with Human Feedback). However, it is important to remember that human annotators may also exhibit biases in their annotations. Indeed, we have some evidence of this in our Prolific annotation data, where we observed that male annotators were 25 percentage points less likely than female annotators to report that a female appellant was the property owner (p = 0.015), and male annotators were also 10 pp more likely than female annotators to report that a male appellant was the property owner (p = 0.135). Our point here is that human annotators are also very likely to exhibit gender bias in their annotations. Thus, bias is a challenge with annotation, broadly, not specifically with annotation employing M-LLMs.

\vspace{2mm}

\textbf{Approach:} We compared Gemini’s responses to annotations from two different sets of human evaluators: (i) independent responses from Prolific workers and (ii) consensus-based coding from a group of high school students. This analysis provides insight into the extent to which generative AI can serve as a reliable tool for extracting behavioral and tonal features from unstructured audio data. Human annotations were collected from two samples, in two ways:

\begin{enumerate}

\item \textit{Prolific Workers (Independent Coding):} Several hundred human coders were recruited through Prolific to independently annotate a random sample of 75 audio recordings. Multiple coders evaluated each recording. We attempted to assign workers to recordings at random. However, the randomization approach, based on a random selection of audio file indices using a random integer draw, resulted in a great deal of imbalance in the assignment of records to workers. Further, we omitted annotations from Prolific workers who reported that the audio clarity was below 75\%, or where the annotator completed the annotation task too quickly (i.e., where the time to task completion was less than the duration of the audio files). Accordingly, we obtained usable annotations for only 59 recordings and we obtained a much larger number of annotations for some recordings than others. Ultimately, we obtained responses from multiple annotators for only 42 of the recordings. 

Each Prolific worker was navigated to a Qualtrics survey, which embedded the relevant hearing audio files, as well as survey items to capture annotations. Workers were first presented with a series of instructions. They were told that they would need to annotate two audio recordings, in sequence. They were also informed of the context of the audio files (i.e., that each file was a recording of a property tax protest hearing that had occurred in Houston, Texas). Workers were instructed to listen to the audio file in its entirety before submitting their responses, and they were also provided with a free-form essay-style text box for note-taking. To help guide the workers in their annotations, we provided an example of some notes that we had taken previously when undertaking pilot coding of some hearing audio files. Thus, the Prolific workers had some examples of the types of things they should pay attention to when listening to the audio files, to be able to answer correctly. Workers responded to a series of survey items that mirrored a subset of the items to which Gemini responded. We consider only a subset of items to shorten the workers' tasks and thereby keep costs down. Nonetheless, the median time to task completion was 35 minutes, including time spent listening to the audio recordings, which ranged from 209 seconds to 1,956 seconds; that is, the shortest hearing record was a little more than 3 minutes long, whereas the longest lasted more than 30 minutes. In total, this annotation exercise cost slightly more than \$2,500 to conduct. 

Using the resulting annotation data, we conduct two types of analyses. First, for each question-recording pair, we obtain a measure of inter-coder agreement, to obtain a sense of the subjectivity inherent in each question. Second, for each question, we calculate the pairwise agreement between Gemini and `Prolific Coders', generally. We achieve this by stacking all Prolific-worker responses to a given question, across recordings and then comparing those responses with Gemini's responses to the same question-recording pair, obtaining a single measure of pairwise agreement.   

\item \textit{Student Coders (Consensus-Based Coding):} A random sample of 50 audio recordings was annotated by three high school students, all based in Dallas, Texas. The students were unpaid, however, they participated because they were seeking to obtain research experience with faculty. The student annotators were thus highly motivated, and likely more attentive than Prolific workers in their annotation efforts. At the outset, the students completed annotations of 2 recordings and met with one of the co-authors to discuss their annotations and ensure they correctly understood the task. The students took approximately 3 months to complete their annotations of 50 recordings, in large part because their annotation efforts required a great deal of discussion and repeated listening. The students were similarly informed about the context of the audio recordings they would hear. They were asked to listen to each audio recording together and to discuss what they heard to arrive at a consensus about each answer. The students met together on Zoom to perform the annotations, listening to the audio together and pausing every few minutes to discuss what they had heard. Ultimately, they recorded their annotations in a spreadsheet. We ultimately obtained a single single set of annotations from these students, per recording, which we use to conduct a single analysis based on a pairwise comparison of inter-coder agreement across recordings. 

\end{enumerate}

\vspace{2mm}

\textbf{Measures \& Results:} We begin by presenting raw correlations between Prolific workers' annotations and Gemini annotations for various items in Table~\ref{tab:gemini_prolific_corr}. The results are rank-ordered, with items exhibiting the strongest agreement appearing at the top. The first point of observation is that we observe positive correlations for every item. That said, we also observe a great deal of variation in agreement across the items. Gemini agrees with Prolific annotators very often when it comes to an appellant's gender, whether the appellant is the property owner, whether the hearing took place online, and whether the appellant presented any formal evidence. However, there is relatively little agreement for some items, such as whether an appellant sounds nervous, sounds defensive, or raised their voice.

\begin{table}[h]
    \centering
    \footnotesize
    \caption{Pairwise Correlations Between Gemini and Prolific Annotations}
    \label{tab:gemini_prolific_corr}
    \begin{tabular}{l c}
        \toprule
        \textbf{Annotation Item} & \textbf{Correlation} \\
        \midrule
        Appellant is Female & 0.678 \\
        Appellant is the Property Owner & 0.409 \\
        Hearing Took Place Online & 0.336 \\
        Appellant Presented Formal Evidence & 0.291 \\
        Appellant Asked Questions & 0.253 \\
        Appellant Asked Someone to Speak Up & 0.241 \\
        Appellant Sounded Frustrated & 0.224 \\
        Appellant Ever Disagreed with Someone & 0.192 \\
        Appellant Ever Interrupted Someone & 0.181 \\
        Appellant Sounded Confident & 0.153 \\
        Appellant Asked Someone to Repeat Themselves & 0.145 \\
        Appellant Ever Raised Their Voice & 0.075 \\
        Appellant Sounded Defensive & 0.036 \\
        Appellant Sounded Nervous & 0.015 \\
        \bottomrule
    \end{tabular}
\end{table}

Although the above analysis provides a high-level indication of associations and already suggests that Gemini is indeed providing useful information in its annotations, we next turn to a more rigorous analysis. To quantify the level of agreement between Gemini and human annotators or among the human annotators alone, for a particular question/item, we employ two measures: Cohen's Kappa and Krippendorff's Alpha, respectively. These are two established measures of inter-coder agreement \citep{cohen1960coefficient,krippendorff2011computing}. Cohen's Kappa quantifies agreement between pairs of annotators assuming no missingness (i.e., each coder responds with an annotation about each instance), whereas Krippendorff's Alpha can be used to quantify agreement among three or more coders, allowing for missing values (i.e., non-response from some coders in some cases). The results of these analyses are summarized in Table~\ref{tab:gemini_validation}.  

We observe that differences emerge depending on which group of annotators we consider. The high school students, who worked together and reached consensus, exhibit higher agreement with Gemini on some dimensions than the Prolific workers, and vice versa. For example, Gemini exhibits greater agreement with the Prolific workers when it comes to annotating whether the appellant has presented evidence at their hearing, whether the appellant asks questions, or whether the appellant verbally disagrees with another party. By contrast, Gemini is more likely to exhibit agreement with the high school students when annotating whether the appellant has raised their voice or sounds frustrated. When we consider the average agreement among annotators for different questions (across hearings), we see that the pattern of agreement is not necessarily dependent on whether an annotation is characterized by inherent subjectivity. For example, some behaviors that elicit strong annotation agreement among Prolific workers, e.g., the presentation of evidence by an appellant, whether the appellant asks another party to speak up or repeat themselves, whether the appellant interrupts another party, and whether the appellant raises their voice, do not necessarily exhibit stronger agreement with Gemini's annotations. 

\clearpage
\begin{landscape}
\begin{table}[h]
    \centering
    \footnotesize
    \caption{Agreement Levels Between Gemini and Human Coders}
    \label{tab:gemini_validation}
    \begin{tabular}{l l p{4cm} p{4cm} p{4cm}}  % Set column widths
    \toprule
    Test Statistic & & \textbf{Cohen's Kappa} & \textbf{Cohen's Kappa} & \textbf{Krippendorff's Alpha} \\
        \midrule
    \textbf{Level} & \textbf{Value} & \multicolumn{1}{p{4cm}}{\raggedright \textbf{Agreement Between Gemini \& Prolific Coders}} & \multicolumn{1}{p{4cm}}{\raggedright \textbf{Agreement Between Gemini \& High School Coders}} & \multicolumn{1}{p{4cm}}{\raggedright \textbf{Mean Agreement Between Prolific Coders Across Recordings}} \\
    \midrule
        \textbf{Near Perfect} & 0.81 – 1.00 & Gender\textsuperscript{***} & Asks to Speak Up\textsuperscript{n/a} & Gender\textsuperscript{***} \\
        \textbf{Substantial} & 0.61 – 0.80 & -- & Gender\textsuperscript{***}, Is Owner\textsuperscript{***} & -- \\
        \textbf{Moderate} & 0.41 – 0.60 & Shows Evidence\textsuperscript{***} & Online Hearing\textsuperscript{***} & Shows Evidence\textsuperscript{***}, Is Owner\textsuperscript{***}, Asks to Speak Up\textsuperscript{***}, Interrupts\textsuperscript{***}, Raises Voice\textsuperscript{***} \\
        \textbf{Fair} & 0.21 – 0.40 & Online Hearing\textsuperscript{*}, Is Owner\textsuperscript{*}, Asks Questions\textsuperscript{*}, Verbally Disagrees\textsuperscript{*} & Raises Voice\textsuperscript{*}, Sounds Confident\textsuperscript{*}, Sounds Defensive\textsuperscript{*} & Online Hearing\textsuperscript{*}, Asks to Repeat\textsuperscript{*}, Sounds Confident\textsuperscript{*}, Sounds Frustrated\textsuperscript{*} \\
        \textbf{Slight} & 0.01 – 0.20 & Asks to Repeat, Asks to Speak Up, Raises Voice, Sounds Confident & Sounds Frustrated\textsuperscript{+}, Interrupts, Asks Questions, Verbally Disagrees, Shows Evidence & Asks Questions, Verbally Disagrees, Sounds Defensive \\
        \textbf{None} & < 0.01 & Interrupts, Sounds Frustrated, Sounds Defensive, Sounds Nervous\textsuperscript{n/a} & Asks to Repeat, Sounds Nervous\textsuperscript{n/a} & Sounds Nervous \\
        \bottomrule
    \end{tabular}
    \begin{flushleft}
        \textit{Note}: + $p < 0.10$; * $p < 0.05$; *** $p < 0.001$. "n/a" indicates cases where statistical significance is not obtainable as the test statistic is undefined due to a complete lack of variation in responses from either Gemini or human coders; Asks to Speak Up is recorded as near perfect agreement between Gemini and High School coders because, although there is no variation in annotations among the High School students (they respond no in every case), Gemini almost always agrees; Sounds Nervous is recorded as yielding no evidence of agreement between Gemini and the high school students because Gemini exhibits no variation in its response across recordings, i.e., it always answers no, yet the High School students respond Yes in nearly 50\% of cases.
    \end{flushleft}
\end{table}
\end{landscape}

Given these results, one broad conclusion we draw is that Gemini's annotations do contain meaningful information; they are not merely noise. For example, when it comes to annotating an appellant's gender, whether the appellant is the property owner, whether the hearing has taken place online, and to a lesser extent whether the appellant sounds confident or defensive, Gemini consistently exhibits statistically significant agreement with human annotators. 

\clearpage
\section*{APPENDIX C: Explanation of Tones and Behaviors Coded in Audio Recordings}

The following descriptions define the tones and behaviors coded from audio recordings of property tax protest hearings in Harris County, Texas. These features represent key elements of appellant communication and interaction during hearings. The coding was performed using Gemini 1.5 Flash and reflects distinct and interpretable aspects of speech and behavior that are generally comprehensive in the context of such hearings.

\begin{itemize}
    % Tones
    \item \textbf{Confidence}: Confidence is characterized by a steady, clear, and self-assured vocal delivery, often marked by a lack of hesitation and a consistent tone. Projecting confidence can instill a sense of credibility and competence in listeners.

    \item \textbf{Anger}: Anger is characterized by feelings of irritation, frustration, or rage. It is typically a temporary response to a perceived threat, injustice, or frustration. Anger is identified by a strained voice, faster or slower speech, sharper or clipped word pronunciation, and a lack of warmth or softness in tone

    \item \textbf{Hostility}: Hostility is identified through a sharp, cutting tone, often paired with elevated volume, sarcasm, or aggression. This behavior may alienate listeners.

    \item \textbf{Frustration}: Frustration is conveyed through strained or uneven vocal tones, frequent sighs, or a quicker-than-usual pace, signaling dissatisfaction or impatience with the proceedings.

    \item \textbf{Annoyance}: Annoyance is a less intense and more transient form of frustration, often marked by clipped or curt vocal delivery, with signs of irritation but not outright aggression.

    \item \textbf{Defensiveness}: Defensiveness is indicated by a protective or guarded tone, characterized by justifications or explanations in response to perceived criticism.

    \item \textbf{Nervousness}: Nervousness is detected through hesitations, voice tremors, pitch variability, and speech errors, reflecting discomfort or lack of confidence.

    % Behaviors
    \item \textbf{Disagrees}: This behavior captures instances where one party verbally expresses disagreement with another party's statements, either explicitly or implicitly. Disagreement is important as it can shape the dynamics of argumentative discourse.

    \item \textbf{Presents Evidence}: This behavior indicates instances where the appellant references, describes, or provides documentation, photos, spreadsheets, etc., in support of their argument. Presenting evidence is a critical behavior in structured hearings, as it demonstrates preparation and engagement.

    \item \textbf{Raises Voice}: This behavior captures instances of increased vocal volume, which may signal emphasis, frustration, or attempts to assert control in the interaction.

    \item \textbf{Asks Questions}: This includes any inquiries from one party, e.g., the appellant, to another, e.g., the panel, reflecting an attempt to seek clarification or challenge the hearing process. Question-asking is a critical component of participatory dialogue.

    \item \textbf{Interrupts}: This behavior reflects overlaps in speech where one party interjects before another has finished speaking. Interruptions may indicate urgency, frustration, or a desire to assert control.

    \item \textbf{Asks to Speak Louder}: This behavior reflects difficulties in hearing or understanding another party, prompting requests for increased vocal volume.

    \item \textbf{Asks to Repeat}: This behavior reflects instances where one party requests another to restate their points, e.g., due to inaudibility or confusion.
\end{itemize}

These coded tones and behaviors were selected for their relevance to hearing dynamics and their grounding in existing communication and behavioral research. They aim to comprehensively capture the range of appellant interaction styles, contributing to the analysis of procedural outcomes and dynamics.

%\clearpage
%\appendix
%\input{AEJ/paper-appendix.tex}

\end{document}